\documentclass[a4paper,fleqn]{cas-sc}

\usepackage{include}

\def\tsc#1{\csdef{#1}{\textsc{\lowercase{#1}}\xspace}}
\tsc{WGM}
\tsc{QE}
\tsc{EP}
\tsc{PMS}
\tsc{BEC}
\tsc{DE}

\begin{document}

\let\WriteBookmarks\relax
\def\floatpagepagefraction{1}
\def\textpagefraction{.001}
\shorttitle{Microsimulation of Energy and Flow Effects from Optimal Automated Driving in Mixed Traffic}
\shortauthors{T. Ard et~al.}

\title[mode = title]{Microsimulation of Energy and Flow Effects from Optimal Automated Driving in Mixed Traffic} % AD suggestion (two lines)

\author[1]{Tyler Ard}[type=editor,
                    %auid=000,bioid=1,
                    orcid=0000-0002-5123-1226
                    ]
\cormark[1]
\ead{trard@clemson.edu}

\credit{Methodology, Software, Data curation, Formal analysis, Investigation, Writing - original draft}
\author[1]{Robert Austin Dollar}[%type=editor,
                    %auid=000,bioid=1,
                    orcid=0000-0002-2618-3066
                    ]

\ead{rdollar@clemson.edu}

\credit{Methodology, Software, Writing - review \& editing}
\author[1]{Ardalan Vahidi}[%type=editor,
                    %auid=000,bioid=1,
                    orcid=0000-0002-1669-3345
                    ]

\ead{avahidi@clemson.edu}

\credit{Conceptualization, Project administration, Supervision, Funding acquisition, Writing - review \& editing}
\author[2]{Yaozhong Zhang}[%type=editor,
                    %auid=000,bioid=1,
                    % orcid=0000-0001-7511-2910
                    ]

\ead{yaozhong.zhang@anl.gov}

\credit{Data curation, Validation, Writing - review \& editing}
\author[2]{Dominik Karbowski}[%type=editor,
                    %auid=000,bioid=1,
                    % orcid=0000-0001-7511-2910
                    ]

\ead{dkarbowski@anl.gov}

\credit{Data curation, Validation, Writing - review \& editing}
\cortext[cor1]{Corresponding author}

\address[1]{Mechanical Engineering, Clemson University, Clemson, SC 29634}
\address[2]{Argonne National Laboratory, 9700 S Cass Ave, Lemont, IL 60439}

\begin{abstract}
    This paper studies the energy and traffic impact of a proposed Anticipative Cruise Controller in a PTV VISSIM microsimulation environment. 
    We dissect our controller into two parts: 1. the anticipative mode, more immediately beneficial when automated vehicle fleet penetration is low, and 2. the connected mode, beneficial in coordinated car-following scenarios and high automated vehicle penetrations appropriate for autonomous vehicle specific applications. Probabilistic constraints handle safety considerations, and vehicle constraints for acceleration capabilities are expressed through the use of powertrain maps.
    Real traffic scenarios are then modeled using time headway distributions from traffic data. To study impact over a range of demands, we vary input vehicle volume from low to high and then vary automated vehicle penetration from low to high.
    When examining all-human driving scenarios, network capacity failed to meet demand in high-volume scenarios, such as rush-hour traffic. We further find that with automated vehicles introduced which utilize probabilistic constraints to balance safety and traffic compactness, network capacity was improved to support the high-volume scenarios.
    Finally, we examine energy efficiencies of the fleet for conventional, electric, and hybrid vehicles. We find that automated vehicles perform at a 10\% - 20\% higher energy efficiency over human drivers when considering conventional powertrains, and find that automated vehicles perform at a 3\% - 9\% higher energy efficiency over human drivers when considering electric and hybrid powertrains. Due to secondary effects of smoothing traffic flow, energy benefits also apply to human-driven vehicles that interact with automated ones. Such simulated humans were found to drive up to 10\% more energy-efficiently than they did in the baseline all-human scenario.
\end{abstract}

\begin{graphicalabstract}
\includegraphics[page=1]{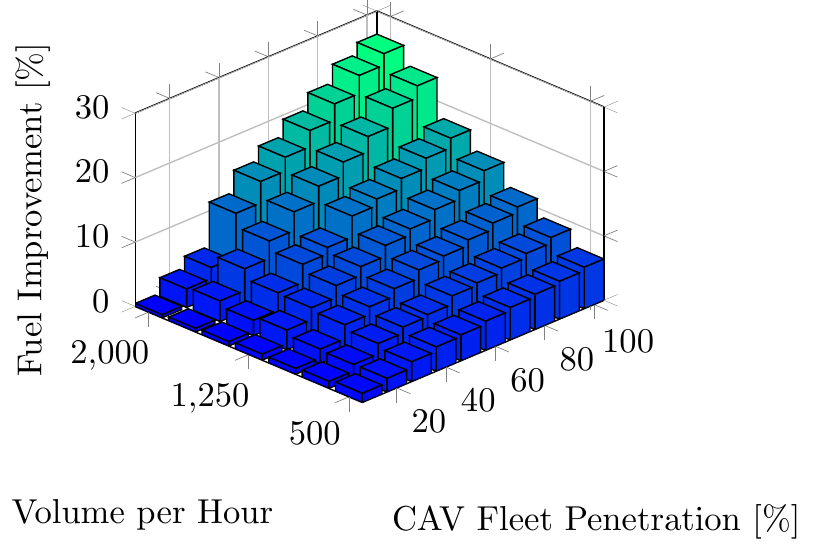} 
\end{graphicalabstract}

\begin{highlights}
\item Considers conventional, electric, and hybrid vehicle types for energy effects.
\item Introduce probabilistic constraints for safety and traffic flow considerations.
\item Integrates empirical data and high fidelity models for increased realism in simulation.
\item Mixed fleets observe improved energy and flow effects.
\end{highlights}


\maketitle

%%%%%%%%%%%%%%%%%%%%%%%%%%%%%%%%%%%%%%%%%%%
%%% Introduction                        %%%
%%%%%%%%%%%%%%%%%%%%%%%%%%%%%%%%%%%%%%%%%%%

\section{Introduction} \label{Introduction}

Human drivers are often reactive when following other cars, as their view is often blocked by the preceding car and therefore their event horizon is limited. In sudden slowdowns, they often fail to consider their impact on upstream traffic. This is not only disruptive to traffic flow and is unsafe, but it can result in inefficient slow-down of multiple vehicles. Balancing the position dynamically with respect to the cars in the front and back is cognitively demanding for humans. Most autonomous cars without connectivity do not necessarily do better: many are designed to behave like human-driven vehicles and could be reactive to the perception of their immediate surroundings, which results in similar short-sighted decisions. At the same time, autonomous vehicles offer unprecedented  potentials for boosting road safety, capacity, and efficiency \citep{vahidi2018energy}. This is because of their ability to process data from many more sources (e.g. Vehicle-to-Everything (V2X) fused with on-board sensing) and perform much more precise positioning and control than human drivers could.  While similar information can be processed, and provided to connected human-driven vehicles, only automated vehicles can be made to comply with, and reliably follow, real-time energy-efficient commands.

\begin{figure}
    \centering
    \input{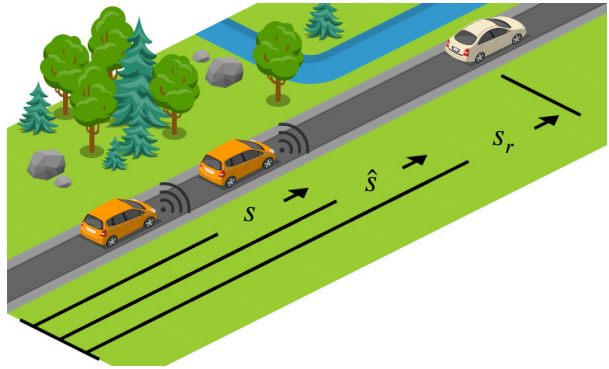}
    \caption{Connected vehicles (orange) following in a string with an unconnected, human-driven vehicle (white). Control, output, and reference states are shown. This figure was generated at \url{https://icograms.com}.}
    \label{fig:Icograms_CAV}
\end{figure}

Automated Driving Assistance Systems (ADAS) such as Adaptive Cruise Control (ACC) and Cooperative Adaptive Cruise Control (CACC) have been proposed for car-following and platooning, such as those depicted in Figure \ref{fig:Icograms_CAV}, which can rely on instantaneous information about the vehicle surroundings and therefore do not engage proactively. Here, ACC vehicles decide their velocity based on traffic information from sensory data, whereas CACC vehicles additionally utilize information broadcast to each other to make more informed decisions but still are not anticipative in nature. Several of such commerical automated longitudinal control systems have been shown to be string unstable, leading to inefficient road utilization and propagation of traffic disturbances \citep{Gunter2019}. 

By better anticipating the imminent motion of the preceding vehicle, via Vehicle-to-Vehicle (V2V) connectivity or by systematic probabilistic reasoning, an automated vehicle can drive more smoothly - contributing to energy efficiency and traffic flow. This paper studies the traffic flow and energy impact of optimal anticipative cruise control among heterogeneous fleets of human and automated passenger vehicles via large scale microsimulations. The control algorithm itself builds on the receding horizon motion planning proposed in \citep{Dollar2018EfficientStrings}, with the new additions of chance constraints and time headway tracking to improve performance when following human drivers. There have been previously proposed optimal controllers explicitly designed to improve energy efficiency and traffic flow \citep{kamal2011ecological, kamal2014smart}, though the microsimulation implementation we take allows observation of emergent behavior in interactions of a large mix of human driven and anticipative automated vehicles, and thus provides new insights on system-wide energy and flow impacts. To calculate the energy impact most realistically, velocity profiles of all micro-simulated vehicles are tracked in high fidelity vehicle models from Autonomie \citep{autonomie}. Therein, we separately consider fleets of gasoline engine, battery electric, and hybrid electric vehicles, and we report the energy impact from optimal automated driving for each class of vehicles. 

Some literature on car-following and platooning has conducted fuel economy studies using drive cycle data to impose traffic conditions. However, various standard drive cycles have been shown to overestimate fuel economies of vehicles for similar intensities of traffic conditions in studies using real driving data \citep{RykowskiFuelEconomy}. In contrast, \citet{Hou2015SuitabilityAnalysis} show that current microsimulation software is capable of producing fuel estimations that approximately match those generated from real data. Furthermore, it can also be shown that ACC and CACC vehicles can smooth traffic and prevent disturbances from growing. So, imposing fixed drive cycles present artificial conditions: fixed drive cycles impose the same traffic disturbances, despite changes in fleet behavior, and may not represent the full impact of automation. A microsimulation approach can realize traffic conditions in an organic manner - particularly given that the parameters used produce simulation data similar to real data. By replicating real-world traffic conditions in a microsimulation, a study can then predict traffic impact due to changes in control algorithms used \citep{Hollander2008TheModels}. Compared to drive cycle studies like \citep{Dollar2018EfficientStrings, stanger2013model, wan2017probabilistic}, the use of microsimulation software enables more realistic performance assessment. 

We use PTV VISSIM 10 \citep{ptvvissim10} for microsimulations. VISSIM employs the Wiedemann car-following model \citep{WIE74}, which is a psycho-physical driver model for human driving. We have been able to custom-code our anticipative car-following algorithms for the automated vehicle class via a Dynamic Link Library (DLL) that VISSIM provides. Additional control of traffic is permitted through stochastic distributions of vehicle flux into a network and preferred acceleration or deceleration behavior, and altered driving behavior to account for road-grade and controlled intersections to contribute to easy real-world modelling \citep{WashingtonStateDepartmentofTransportationWSDOT2014ProtocolSimulation}. It has been shown that VISSIM simulation results are sensitive to simulation parameters, but only small changes to default parameters are needed to replicate observed traffic phenomena well - it is sufficient to alter headway parameters alone \citep{Hou2015SuitabilityAnalysis, Gomes2004CongestedVISSIM, Dong2015VISSIMFreeways, FellendorfVISSIM}. From this, microsimulation software can be used to well-realize traffic dynamics and thus predict benefits of a control strategy. Fuel economy improvement has already been shown for traffic controllers handling ramp access and intersections \citep{Tettamanti2008ModelManagement, Tettamanti2012DevelopmentEnvironment, CassandrasVissim}. VISSIM was also used in \citep{Yu2014AnRisk} to evaluate safety improvements from a variable speed limit controller. Finally, \citet{LiuFreeway} offer a rule-based connected car-following and lane-change controller to study energy and traffic effects in a fashion similar to this work, though we aim to isolate and study car-following more explicitly.

While experiments like \citep{Gunter2019, lidstrom2012modular, ploeg2011design} have the advantage of using real vehicles, they typically involve isolated strings of fewer than 10 of them. The larger scale studies, for instance \citep{stern2018dissipation} with 22 vehicles in a loop formation, are interesting but do not quite emulate the complex scenarios that arise in longer linear queues. The microsimulation platform proposed here complements, but does not replace, experiments in favor of larger-scale evaluation.

In the rest of this paper, first the optimal control formulation of the anticipative cruise control algorithm is proposed in Section \ref{sec:Optimal Control Formulation}, which by penalizing acceleration of the vehicle helps energy efficiency and has secondary benefits due to traffic harmonization. Traffic compactness is considered through the use of headway tracking terms in the objective, and further handled as a trade-off between vehicle safety and gap as a probabilistic constraint. Then, the setup for a VISSIM environment is depicted and simulation parameters are carefully justified in Section \ref{sec:VISSIM Setup}, where real-scenario conditions are replicated. The process to customize control of automated vehicles in VISSIM is also described. The energy efficiency of strings of vehicles are then evaluated in high-fidelity powertrain software Autonomie, and benefits to traffic flow and discussion of its effects to energy use for both human and automated vehicles are finalized in Section \ref{sec:Simulation Results}. Section \ref{sec:Conclusion} concludes the paper with summary of results and directions for future work. 

%%%%%%%%%%%%%%%%%%%%%%%%%%%%%%%%%%%%%%%%%%%
%%% Optimal Control Formulation         %%%
%%%%%%%%%%%%%%%%%%%%%%%%%%%%%%%%%%%%%%%%%%%
\section{Optimal Control Formulation}\label{sec:Optimal Control Formulation}
The longitudinal control of the automated vehicles is formulated in a Model Predictive Control (MPC) framework. As detailed further in the following subsection, a finite sequence of future acceleration commands is decided by minimizing a control objective function subject to the longitudinal dynamics, actuator, and safety constraints. Only the first control in this sequence is applied at each step in time, the optimization window recedes forward, and the optimization is resolved. Optimizing a sequence of control moves has the advantage of planning for the near future, so that immediate decisions made are beneficial for later, predicted scenarios. The receding horizon implementation introduces feedback and realizes closed-loop control \citep{MaciejowskiPredictive}. Due to the linearity of the model and the constraints, and because we choose a quadratic function of the control input and states as the objective, the optimization problem is converted to a quadratic program (QP) for which efficient numerical solution methods exist.

An overview of the optimal control formulation for car-following is then given as follows. A linear set of discrete equations replicate the simulation model used in VISSIM, a quadratic objective is proposed to smooth acceleration experienced by the MPC-controlled vehicle and maintain a reference time headway from the preceding vehicle, and constraints are formulated to set vehicle capabilities and impose safe driving. Further details, such as an optimal terminal constraint derivation based on particle kinematics, can be found in \citep{Dollar2018EfficientStrings}.

%%%%%%%%%%%%%%%%%%%%%%%%%%%%%%%%%%%%%%%%%%%
\subsection{Control-Oriented Longitudinal Dynamics Model}
We consider the longitudinal dynamics of a given MPC-controlled vehicle (ego) to be the kinematic relations between commanded acceleration and position. In discretization, the Euler integration method is chosen to match the simulation method VISSIM employs. As depicted in Figure \ref{fig:Icograms_CAV}, let $s, v, a$ be the absolute position, velocity, and realized acceleration of the ego, let $u$ be the commanded acceleration of the ego, and let $i$ denote the stage in the control horizon. Then, the control-oriented model has the following state space form.
\begin{equation}\label{eq:model}
\begin{split}
    &\begin{bmatrix}
    s \\ v \\ a
    \end{bmatrix}_{\;i+1}
    =
    \underbrace{
    \begin{bmatrix} 
    1 & \Delta t & 0.25\Delta t^2 \\
    0 & 1 & 0.5\Delta t \\
    0 & 0 & 0
    \end{bmatrix}
    }_{\displaystyle A_d}
    \begin{bmatrix}
    s \\ v \\ a
    \end{bmatrix}_{\;i}
    +
    \underbrace{
    \begin{bmatrix} 
    0.25\Delta t^2 \\
    0.5\Delta t \\
    1
    \end{bmatrix}
    }_{\displaystyle B_d}
    u_i \\
    &\begin{bmatrix}
    \hat{s} \\ a
    \end{bmatrix}_i
    =
    \underbrace{
    \begin{bmatrix} 
    1 & T_H & 0 \\
    0 & 0 & 1
    \end{bmatrix}
    }_{\displaystyle C_d}
    \begin{bmatrix}
    s \\ v \\ a
    \end{bmatrix}_{\;i}
\end{split}
\end{equation}
Here, subscript $d$ denotes the discrete version of the control matrix, $\Delta t$ is the MPC discretization step size. The desired time headway between the ego and its preceding vehicle is denoted by $T_H$ and defines the output variable $\hat{s}=s+T_Hv$. The latter, which linearly increases with ego speed, is used when penalizing deviation from a desired headway in the next subsection. We also choose to include ego's acceleration as an output since it will be penalized in the control objective function.

%%%%%%%%%%%%%%%%%%%%%%%%%%%%%%%%%%%%%%%%%%%
\subsection{Objective Function}\label{sec:objective}
With the purpose of boosting fuel economy of the ego vehicle, a cost function may choose to directly penalize fuel consumption. This poses challenges when solving optimization routines online, however, due to the highly nonlinear nature of such a computation. So, the optimizing routines can occasionally fail to converge to a suitable optimality tolerance, fail to generate a feasible solution, or suffer large, inconsistent runtimes. In a vehicle driving application, these qualities can compromise safety and effectiveness.

With these considerations, a quadratic cost function can be chosen to reliably generate efficient, fast solutions when paired with a linear model. To promote velocity smoothing and eco-driving, as well as promote traffic compactness and safe driving, we choose cost function $J$ to penalize ego acceleration and ego deviation from the time headway - using weights $q_a$ and $q_d$, respectively.
\begin{equation} \label{eq:cost}
    J = q_d[\hat{s}(N) - (s_r(N) - d_r)]^2 + q_a a^2(N)
      + \sum_{i=0}^{N-1} \Big[q_d[\hat{s}(i) - (s_r(i) - d_r)]^2 + q_a[a^2(i) + u^2(i)] \Big]
\end{equation}
\noindent Here, $N$ is the optimization horizon length, $d_r$ is the desired gap at stand-still, and $s_r$ is the preceding vehicle (PV) position. Because the future trajectory of $s_r$ is unknown, further considerations need to be made - as further detailed in Section \ref{sec:Prediction}. 

Clearly, this cost function does not directly minimize fuel consumption. However, penalizing acceleration cuts wasteful braking events, so the above objective aids in reducing energy usage in a computationally efficient quadratic manner. Acceleration minimization also directly attenuates speed disturbances, thus bringing about secondary traffic smoothing benefits which may have secondary benefits in lowering energy consumption of following vehicles. This result is further shown in Section \ref{sec:FuelEconomy}.

%%%%%%%%%%%%%%%%%%%%%%%%%%%%%%%%%%%%%%%%%%%
\subsection{Prediction}\label{sec:Prediction}
To generate a trajectory of $s_r$, we consider two cases: connected driving, and unconnected driving.

In connected driving, the preceding automated vehicle broadcasts its planned intentions as generated from its own optimization routine. Here, solely the future positions of the PV are considered in the optimization routine.

Likewise, in the unconnected case, anticipation of future positions of the PV is handled through the use of a prediction model. For this work, a constant acceleration model was assumed, where acceleration of the PV is propagated forward until velocity saturation at $v_{min}$ or $v_{max}$. Here, we consider the recursive equations as given by Equation \eqref{eq:prediction} for $i \in 0,...,N-1$.
\begin{subequations}\label{eq:prediction}
\begin{align}
        &v_r(i+1) = \min{\left(\max{\left(v_r(i) + a_r(0)\Delta t, \; v_{min}\right)}, \; v_{max}\right)} \\
        &s_r(i+1) = s_r(i) + v_r(i)\Delta t + \mfrac{1}{2}a_r(0)\Delta t^2
\end{align}
\end{subequations}
Other solutions for prediction exist, such as constant velocity modeling \citep{mcdonough2013stochastic}, Markov modeling \citep{mcdonough2014stochastic, zhangPredictive2011}, etc. At first, a constant velocity model was evaluated, but slightly worse performance in fuel economy was observed. More importantly, a much higher risk for collisions and worse traffic compactness were also found. 

%%%%%%%%%%%%%%%%%%%%%%%%%%%%%%%%%%%%%%%%%%%
\subsection{Constraint Handling}
 
\subsubsection{Vehicle Capabilities}
\begin{figure}
    \centering
    \hspace{-2em}
    \input{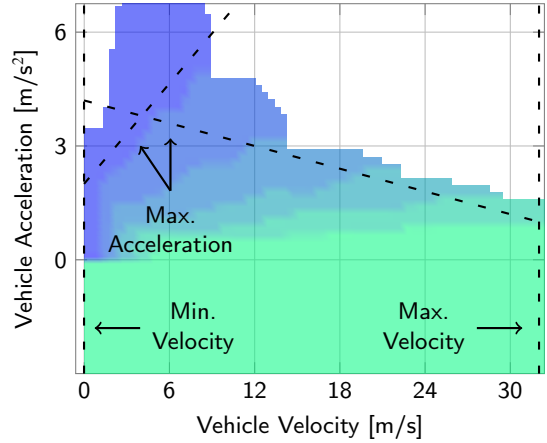}
    \caption{Constraints of the ego with regards to acceleration capabilities at a given speed and maximum and minimum speed imposed. Colored contours indicate steady-state gear for a given velocity-acceleration pair.}
    \label{fig:constraint}
\end{figure}

Take that, at a given current wheel speed of the ego, a limit on the torque at the wheel exists due to the powertrain. It then follows that a map of acceleration of the vehicle can be derived as a function of vehicle speed and road-grade. To generate a map, we evaluate the resulting vehicle acceleration from a set of applied engine torques through the powertrain and account for losses at the wheel due to friction and aerodynamic drag. In doing so, an assumption of constant road-grade was made. Figure \ref{fig:constraint} depicts such a map, where colored regions depict feasible operating space for the powertrain. 

Maximum vehicle acceleration for a passenger car is then approximated as a convex pair of linear constraints - shown in Figure \ref{fig:constraint}. Then, considering minimum and maximum velocities of the vehicle due to the road laws and further detailed in Section \ref{sec:Scenario}, the convex constraint approximations are given as the following.
\begin{subequations}
\begin{align}
    &-m_1v(i) + u(i) \leq b_1 \\
    &-m_2v(i) + u(i) \leq b_2 \\
    &-v(i) - \epsilon_1(i) \leq -v_{min} \\
    &v(i) - \epsilon_2(i) \leq v_{max}
\end{align}
\end{subequations}
Here, $m$ and $b$ are the slope and intercept of the linear acceleration constraints ($m = \{0.285, -0.121\}$ and $b = \{2.00, 4.83\}$). Additionally, $\epsilon$ is a slack variable, formulated to soften constraints to enable solution to the QP in otherwise infeasible driving routines \citep{MaciejowskiPredictive}. To do so, $\epsilon \geq 0$ is a decision variable included as a linear affine term in the cost function with a significant penalty - e.g. the new cost function becomes the following.
\begin{equation}
J' = J + \sum_{i=0}^N q_\epsilon\bar{\epsilon}(i)
\end{equation}
Here, $q_\epsilon >> q_a$, and the overbar represents a vector of variables \citep{MaciejowskiPredictive}. By this method, the optimizer will relax a constraint with a positive $\epsilon$, if necessary, in order to generate a solution. Because of the large cost of doing so, the controller is still incentivized to drive in a manner that removes the penalty quickly.

\subsubsection{Safety Enforcement}\label{sec:Safety}
As mentioned in Section \ref{sec:Prediction}, anticipation of the future gap positions of the PV is handled through the use of a prediction model in the case of an unconnected vehicle. We propose to enforce the in-horizon and terminal intera-vehicle gap constraints probabilistically with a minimum allowable gap. The following inequality gives the safety constraint.
\begin{equation}\label{eq:pvpos}
    s(i) - \epsilon_0(i) \leq s_{\alpha}(i, \alpha) - d_m
\end{equation}
Here, $d_m$ denotes the minimum allowed following distance, and the safe position $s_{\alpha}(i, \alpha)$ is the anticipated position of the PV with probability $\alpha$ at stage $i$. In other words, the \textit{realized} PV position results in a following distance of at least $d_m$ relative to the planned ego position with probability $\alpha$.

The following probabilistic derivation gives $s_{\alpha}(i, \alpha)$. The first step in this process is to derive the distribution of the PV's position over time. The second step uses this distribution to obtain $s_{\alpha}(i, \alpha)$. 

In the first step, we continue to use the constant acceleration assumption of Section \ref{sec:Prediction} but introduce stochasticity in the acceleration estimate. The dynamics of the stochastic state vector $X_r = \begin{bmatrix} S_r & V_r & A_r \end{bmatrix}^\mathrm{T}$ are modeled as a classic double integrator with state transition matrix $A$, such that $X_r\left(k+1\right) = AX_r\left(k\right)$. Here, state variables in capital letters denote random variables. The PV position $S_r$ and velocity $V_r$ are assumed to be measured with negligible variance at the current time, leaving $X_r(0) = \begin{bmatrix} s_r(0) & v_r(0) & A_r \end{bmatrix}^\mathrm{T}$. After assuming normally distributed acceleration, i.e. $A_r \sim \mathcal{N}\left(a_r, \sigma_A^2\right)$, the initial PV state separates into deterministic and stochastic components, where the random acceleration error $\tilde{A}_r \sim \mathcal{N}\left(0, \sigma_A^2\right)$.
\begin{equation} \label{eq:xrsplit}
    X_r\left(0\right) = \begin{bmatrix} s_r \\ v_r \\ a_r \end{bmatrix}^\mathrm{T} + \begin{bmatrix} 0 \\  0 \\ \tilde{A}_r \end{bmatrix}^\mathrm{T}
\end{equation}
Then, the model propagates forward to step $i$.  At this point, $X_r$ is represented as a combination of its deterministic and stochastic parts $\overline{x}\left(i\right)$ and $\tilde{X}\left(i\right)$, respectively. Notice that since the expectation of acceleration error $E\tilde{A}_r = 0$, $\tilde{X}\left(i\right)$ is entirely zero-mean and the position element of $\overline{x}\left(i\right)$ is the expected position that Section \ref{sec:Prediction} provides.
\begin{equation} 
    X_r\left(i\right) = A^i_d \begin{bmatrix} s_r \\ v_r \\ a_r \end{bmatrix}^\mathrm{T} + A^i_d \begin{bmatrix} 0 \\  0 \\ \tilde{A}_r \end{bmatrix}^\mathrm{T} = \overline{x}_r\left(i\right) + \tilde{X}_r\left(i\right)
\end{equation}
The covariance $\Lambda\left(i\right)$ of $X_r\left(i\right)$ is still needed to characterize $S_r\left(i\right)$. Noticing that $\tilde{X}\left(i\right) \sim \mathcal{N}\left(\mathbf{0}, \ \Lambda\left(i\right)\right)$ and applying aforementioned assumptions yields the following formula.
\begin{equation}
    \Lambda\left(i\right) = A^i_d \Lambda_0 \left(A^i_d\right)^\mathrm{T}, \quad \Lambda_0 = 
    \begin{bmatrix} 0 & 0 & 0 \\
                    0 & 0 & 0 \\
                    0 & 0 & \sigma_A^2 \end{bmatrix}
\end{equation}
In summary, $S_r\left(i\right) = s_r + \tilde{S}_r\left(i\right)$ with $\tilde{S}_r\left(i\right) \sim \mathcal{N}\left(0, \ \sigma_s^2\left(i\right)\right)$ and the variance $\sigma_s^2\left(i\right)$ is extracted from the upper-left corner of $\Lambda\left(i\right)$.

Finally, the cumulative distribution function $F_{\tilde{S}_r \left(i\right)}$ of the position error $\tilde{S}_r$ is inverted to yield $s_{\alpha}(i, \alpha)$, such that probability $P\left(S_r > s_{\alpha}\right) = \alpha$. To make use of the zero-mean $\tilde{S}_r \left(i\right)$, let the distance error $d_{\alpha} = s_{\alpha} - s_{r}$ so that $S_r\left(i\right) > s_{\alpha}\left(i\right)$ and $\tilde{S}_r \left(i\right) > d_{\alpha}$ are equivalent events.  The argument $i$ is omitted in Equations \eqref{eq:invcdfderv1} and \eqref{eq:invcdfderv2} for compactness.
\begin{subequations} \label{eq:invcdfderv1}
\begin{align}
    &\alpha = P\left(S_r > s_{\alpha}\right) = P\left(\tilde{S}_r > d_\alpha\right) \\
    &\alpha = 1 - P\left(\tilde{S}_r \leq d_\alpha\right) = 1 - F_{\tilde{S}_r} \left(d_{\alpha}\right)
\end{align}
\end{subequations}
Considering the symmetry of the zero-mean normal distribution, where $F_{\tilde{S}_r} \left(d_{\alpha}\right) = 1 - F_{\tilde{S}_r} \left(-d_{\alpha}\right)$, the following applies. %$\colon$
\begin{subequations} \label{eq:invcdfderv2}
\begin{align}
    \alpha = 1 - \left(1 - F_{\tilde{S}_r} \left(-d_{\alpha}\right) \right) = F_{\tilde{S}_r} \left(-d_{\alpha}\right) \implies
    d_{\alpha} = - F_{\tilde{S}_r}^{-1} \left(\alpha\right) \implies
    %\boxed{
    s_{\alpha}\left(\alpha\right) = s_{r} - F_{\tilde{S}_r}^{-1} \left(\alpha\right)
    %}
\end{align}
\end{subequations}
Thus, $s_{\alpha}\left(i, \alpha\right)$ is obtained as needed.

\begin{figure}
    \centering
    \hspace*{2em}
    \input{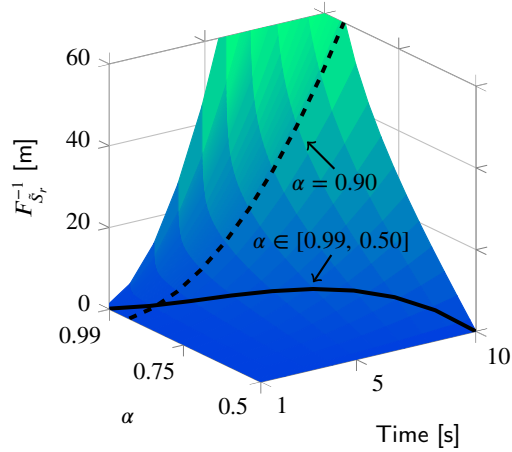}
    \caption{Safety on PV position, $s_{\alpha}$, with the results of the linearly decreasing function $\alpha(i) \in [0.99, 0.50]$, where $\alpha = 0.50$ for time beyond 10 s. Additionally shown is a constant function $\alpha(i) = 0.90$.}
    \label{fig:Salpha}
\end{figure}

A safety probability is then chosen - here, we choose a linearly-decreasing $\alpha(i)$ $\in$ [0.99, 0.50], emphasizing safe driving early in the horizon. Later in the horizon, we hypothesize it is affordable to reduce $\alpha$ in favor of improved traffic compactness and, because of the feedback nature of MPC, safe-driving is still achieved \citep{DOLLARhazardous}.

This is further illustrated in Figure \ref{fig:Salpha}, which depicts the inverse cumulative distribution function of $\tilde{S}_r$ with respect to time and probability. As shown by the dashed line, the increasing uncertainty in PV position forward in time causes large $\alpha$ values later in the horizon to drastically increase the required safe gap, which reduce traffic compactness from the automated vehicles.

%%%%%%%%%%%%%%%%%%%%%%%%%%%%%%%%%%%%%%
\subsection{MPC Calibration Process} \label{sec:tuning}

The list of control parameters used in simulation is given in Table \ref{tab:parameters}. To tune the parameters, VISSIM simulations at 2000 veh/hour/lane were used, as the most demanding traffic scenarios were found under this vehicle flux. This process was done iteratively as follows: 1. Modify $T_H$ to approximately match human drivers and support traffic flow, 2. Modify $q_a$ and $N$ to achieve maximum fuel performance of the fleet of vehicles while meeting optimization time requirements, and 3) Repeat from 1 as necessary to observe benefit in fuel economy. Additionally, $d_m$ was fixed to a low value to promote traffic compactness and approximately match stand-still following distances of human drivers when coming to rest behind another vehicle. In the connected case, $d_r$ was chosen to both promote traffic compactness and operate the vehicles near minimum coefficient of drag conditions due to slipstream effects \citep{watkins2008}, and in the anticipative case, $d_r = d_m$ was chosen so that the controller does not attempt to track a value below the minimum allowable gap when at a stand-still.

It was found that acceleration weighting can be increased considerably in low traffic intensities, which might be exploited to further improve energy efficiency. Instead of using a single set of weights over all traffic demands, more advanced approaches, such as weight scheduling, could modify MPC parameters based on observed local traffic demands. This study uses constant weighting tuned for high traffic density scenarios, leaving weight scheduling for future research.

\begin{table}[]
    \caption{Connected and anticipative algorithm parameters}
    \label{tab:parameters}
    \centering
    \begin{tabular}{c|c|c|l}
        \hline
        Prop. & Anticipative & Connected & Description \\
        \hline
        \hline
        $N_{u}$ & 16 s & 17 s & Horizon length \\
        $q_{a}$ & 2050 & 4000 & Acceleration weight \\
        $q_g$ & 1 & 1 & Gap weight \\
        $q_\epsilon$ & 1e6 & 1e6 & Slack variable weight \\
        $T_{H}$ & 1.4 s & 0.0 s & Time headway \\
        $d_{r}$ & 2 m & 6 m & Distance reference \\ 
        $d_{m}$ & 2 m & 2 m & Minimum distance \\
        \hline
    \end{tabular}
\end{table}

%%%%%%%%%%%%%%%%%%%%%%%%%%%%%%%%%%%%%%%%%%%
\subsection{Quadratic Program Solution and Memory Usage}
To solve the quadratic program as created from MPC, the dual simplex method was interfaced in \texttt{C++} using Gurobi 8 and Eigen 3 \citep{gurobi, eigenweb}. On average, the combined build and solution times for the QP were clocked at 2 ms on a Ryzen 5 1500X processor, whereas the max times reached 10 ms if the slack variables $\epsilon$ were non-zero. Because the controller runs at a sample time of 100 ms, the QP is shown to solve easily online. In addition, peak memory usage for the program was found to be 2 MB.

A plethora of efficient, free QP solvers also exist to ease implementation of the linear MPC, such as CasADi, OSQP, or qpOASES \citep{casadi, osqp, qpoases}.

%%%%%%%%%%%%%%%%%%%%%%%%%%%%%%%%%%%%%%%%%%%
%%% VISSIM Network and Settings         %%%
%%%%%%%%%%%%%%%%%%%%%%%%%%%%%%%%%%%%%%%%%%%
\section{Simulation Environment} \label{sec:VISSIM Setup}
A microsimulation approach models vehicle-to-vehicle interactions, and has the advantage of producing more organic simulations than those in which fixed drive cycles are imposed as driving patterns for leading vehicles. So, replicating real-world traffic conditions can allow a study to predict impact on traffic due to changes in the control algorithms used \citep{Hollander2008TheModels}. 
As such, PTV VISSIM is adopted to model traffic in a realistic manner by modeling human time headways from empirical highway data, as detailed in Section \ref{sec:Wiedemann}. The MPC car-following algorithm is integrated within VISSIM for designated CAVs as described in Section \ref{sec:user-defined}. Furthermore, considerations to isolate the traffic effects due to the optimal controller are made, so: travel times are normalized among simulations, and only flat road is considered. It should be noted that separate optimal controllers and planners can further improve fuel economy over human drivers with these effects considered \citep{LiuLineHaul}. In addition, simulations are run with a timestep of 100 ms \citep{WashingtonStateDepartmentofTransportationWSDOT2014ProtocolSimulation}, while MPC is discretized with a timestep of 1000 ms and run every simulation timestep.

%%%%%%%%%%%%%%%%%%%%%%%%%%%%%%%%%%%%%%%%%%%
\subsection{Human Driver Car-Following Model}\label{sec:Wiedemann}
VISSIM bases their driving model on the Wiedemann (WIE) model for human car-following behavior. In addition, two variations exist: WIE74, suitable for urban and arterial driving, and WIE99, suitable for highway and interstate driving \citep{WIE74, FellendorfVISSIM}. 

The model takes on 4 distinct phases: 1. free-flow, 2. approaching, 3. following, and 4. braking \citep{Lownes2006VISSIM:Analysis}. At steady state car-following conditions, it exhibits limit cycle behavior on the velocity-acceleration phase space, naturally modeling the acceleration oscillation humans tend to exhibit when driving.

\begin{figure}
    \centering
    \input{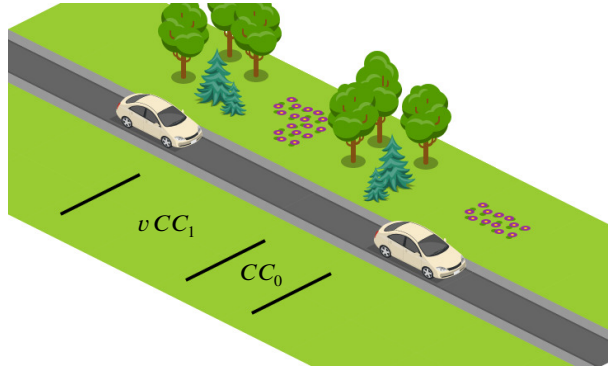}
    \caption{Physical representation of WIE99 headway parameters of time headway and static following distance. This figure was generated at \url{https://icograms.com}.}
    \label{fig:icogramscc}
\end{figure}

To model these phases and transistions from each phase in the WIE model calculations, the WIE99 model has parameters $CC_0, CC_1, CC_2, ..., CC_9$. The physically-intuitive parameters of static following distance, time headway, and following oscillation magnitude are set by $CC_0, \, CC_1, \, \text{and} \,\, CC_2$, respectively, while other model parameters include: following thresholds, acceleration during following oscillation, etc. Figure \ref{fig:icogramscc} depicts the calculated desired following distances employed by WIE vehicles \citep{Dong2015VISSIMFreeways}. 

Guidelines suggest caution when changing VISSIM parameters from the default settings, which are sufficient to use with modifications only to adjusted headway parameters \citep{Dong2015VISSIMFreeways, OregonDepartmentofTransportationODOT2011ProtocolSimulation}. As such, the default parameters are used in all VISSIM simulations \emph{except} for desired time headway, which is replaced with stochastic distributions as found in \citep{Ye2009VehicleData}. This was done because initial testing of high-traffic scenarios showed the resulting time headways of WIE drivers was overly aggressive compared to empirical car-following-car data \citep{Ye2009VehicleData}. Inaccurate results in time headway are especially significant when examining traffic flow with the addition of CAVs, as further detailed in Section \ref{sec:Traffic Flow Results}. Furthermore, \citep{Hou2015SuitabilityAnalysis} calibrate VISSIM by affecting time headway, then show that the fuel economy numbers generated by VISSIM are realistic as compared to empirical traffic data.

\subsection{Customized Car-Following}\label{sec:user-defined}

% \begin{figure}
%     \centering
%     \input{Images/DLLStructure.tex}
%     \caption{Parallel computing architecture for handling multiple vehicles in the network at once on a multi-processor system.}
%     \label{fig:dll structure}
% \end{figure}

To alter driving behavior for vehicles within the network, a separate process is executed in parallel to VISSIM. To accomplish this, a Dynamically-Linked Library (DLL) is attached to VISSIM on startup of a simulation. The DLL has the advantage that its memory is abstracted from VISSIM memory, so computations from both programs can be run with a controlled exchange of data.

On each simulation step, VISSIM checks that a certain vehicle in the network should be externally controlled, and then the calls to and from the DLL are as follows: 1. Ego information such as vehicle ID, velocity, and acceleration are passed to the DLL, as well as preceding vehicle information such as velocity and gap, and 2. Ego commands from the solution of the QP, such as desired acceleration, are retrieved from the DLL and used in dynamic computation at the end of the simulation step. Similar routines are defined for the creation and destruction of a vehicle as it enters and leaves the network, respectively. From this, a parallel computing scheme is defined for all MPC-controlled vehicles.

%%%%%%%%%%%%%%%%%%%%%%%%%%%%%%%%%%%%%%%%%%%
\subsection{Realistic-Scenario Modeling}\label{sec:Scenario}
The goal in creating a VISSIM environment is to reproduce realistic traffic conditions, so that analysis of controller impact on the fleet can then be made \citep{Hollander2008TheModels}. Additionally, it was verified no vehicle collisions occurred throughout the study.

\subsubsection{Transient Considerations}
We first consider that simulations have two sources of unwanted transient effects: 1. Those due to the time period before the network is filled with vehicles, and 2. Those due to the boundary effects in the road network, where once a vehicle leaves the network, those behind it shift to a free-flow driving mode. 

For simulation results discussed in this paper, we only consider simulation time after the first vehicle has left the network, and we define the control volume to span the length of the network with 500m on both ends of the network removed - done to eliminate effects of switching from car-following to free-flow due to the PV leaving the network \citep{WashingtonStateDepartmentofTransportationWSDOT2014ProtocolSimulation, Woody2006CALIBRATINGVISSIM}.

\subsubsection{Physical Location Basis}
%To create the road network for simulation, we consider a 3.65 km subsection of US Route 123-E between Clemson, SC, and Greenville, SC. 
To isolate car-following effects in a realistic scenario, these simulations use the speed limit and 3.65 km length of a real South Carolina road segment between entry and exit ramps. As adapted from \citep{WashingtonStateDepartmentofTransportationWSDOT2014ProtocolSimulation}, human driver maximum velocity is distributed from $+5$ to $-10$ mph of the posted speed limit, minimum velocity is 0, and road grade is neglected.

\subsubsection{Travel Time Management}
It was observed that, due to velocity smoothing effects and dissipation of jams from the autonomous vehicles, average vehicle travel times were improved. To isolate energy saved by the vehicles due to smooth driving, the average travel time of vehicles in the network was normalized, effectively minimizing differences in kinetic action between simulations.

\begin{figure}
    \centering
\begin{tikzpicture}

\begin{axis}[%
width=4.423cm,
height=3.875cm,
at={(0cm,0cm)},
scale only axis,
point meta min=-2.30230136975845,
point meta max=1.13919786451641,
xmin=0,
xmax=100,
xtick={  0,  20,  40,  60,  80, 100},
xlabel style={font=\color{white!15!black}},
xlabel={CAV Fleet Penetration [\%]},
ymin=500,
ymax=2000,
ylabel style={font=\color{white!15!black}},
ylabel={Volume per Hour},
axis background/.style={fill=white},
xmajorgrids,
ymajorgrids,
legend style={legend cell align=left, align=left, draw=black},
colormap={mymap}{[1pt] rgb(0pt)=(0,0,1); rgb(63pt)=(0,1,0.5)},
colorbar,
colorbar style={
    fill opacity=0.75, 
    font=\color{black},
    ylabel={Change [\%]},
    }
]

\addplot[%
surf,
fill opacity=0.75, shader=interp, colormap={mymap}{[1pt] rgb(0pt)=(0,0,1); rgb(63pt)=(0,1,0.5)}, mesh/rows=11]
table[row sep=crcr, point meta=\thisrow{c}] {%
x	y	c\\
0	500	-0\\
0	750	-0\\
0	1000	-0\\
0	1250	-0\\
0	1500	-0\\
0	1750	-0\\
0	2000	-0\\
10	500	-0.114627372242239\\
10	750	-0.171651481113011\\
10	1000	-0.230403078248318\\
10	1250	-0.16616567676077\\
10	1500	-0.542718291741387\\
10	1750	-1.11234841595145\\
10	2000	-1.31231235403737\\
20	500	-0.0105424276846239\\
20	750	-0.164158535813829\\
20	1000	-0.435200748625045\\
20	1250	0.217642766259378\\
20	1500	-0.920101687014554\\
20	1750	-1.71374469706189\\
20	2000	-2.10230136975845\\
30	500	-0.112039001877661\\
30	750	-0.241320177926594\\
30	1000	-0.467494617061579\\
30	1250	0.0319757737937206\\
30	1500	-1.08818068775644\\
30	1750	-1.64786200891352\\
30	2000	-1.90736760716591\\
40	500	-0.14946477608647\\
40	750	-0.283688102180122\\
40	1000	-0.541750443926889\\
40	1250	-0.337074135290174\\
40	1500	-0.510348178353074\\
40	1750	-1.19518865452681\\
40	2000	-0.63664190334785\\
50	500	0.00237660694198003\\
50	750	-0.148579463727528\\
50	1000	-0.492552026417142\\
50	1250	-1.11303057375188\\
50	1500	-0.433899989141218\\
50	1750	-1.13296501232655\\
50	2000	0.300536237981692\\
60	500	0.129770984867249\\
60	750	-0.0316114101753973\\
60	1000	-0.444682701823065\\
60	1250	-0.977134372821979\\
60	1500	0.228185056064535\\
60	1750	-0.572654373104496\\
60	2000	1.13919786451641\\
70	500	-0.0913656906254493\\
70	750	-0.156545537018505\\
70	1000	-0.546106497163274\\
70	1250	-1.07101083175736\\
70	1500	-0.163094768271254\\
70	1750	-0.835774070655008\\
70	2000	0.779683551173689\\
80	500	-0.0691021201512222\\
80	750	-0.147311503776777\\
80	1000	-0.573906486839498\\
80	1250	-1.08049828338031\\
80	1500	0.00254220752058623\\
80	1750	-1.17918088914426\\
80	2000	0.717504764684503\\
90	500	-0.0878062600077528\\
90	750	-0.179544274072621\\
90	1000	-0.603040503407599\\
90	1250	-1.12974868364141\\
90	1500	0.0697671470445288\\
90	1750	-2.36744896431851\\
90	2000	-0.259510616471579\\
100	500	1.86440688365756e-05\\
100	750	-0.0817448596825653\\
100	1000	-0.492075399053114\\
100	1250	-1.06575809485724\\
100	1500	0.169533940914272\\
100	1750	-2.27250794917229\\
100	2000	-0.10732679841472\\
};
% \addlegendentry{data1}

\end{axis}
\end{tikzpicture}%
    \caption{Percentage change in the average travel time of vehicles in the network as compared to the baseline all-WIE scenario of each volume per hour. CAV desired speeds were set to minimize travel time changes.}
    \label{fig:travel_time}
\end{figure}
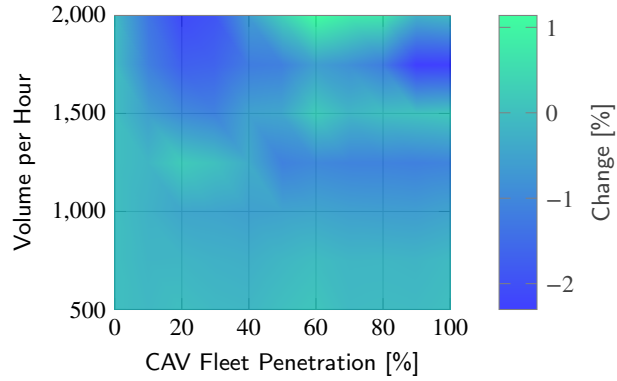

To this end, the maximum velocity of CAVs was tuned such that the average travel times of vehicles in the fleet matched to a tolerance of $\pm2\%$ of the all-WIE scenario with the same input vehicle flux. These results are illustrated in Figure \ref{fig:travel_time}.

To feasibly achieve this result outside of a simulation environment, more advanced approaches could be taken by including a pacing algorithm, such as the analytical high-level planners introduced in \citep{vahidiautomatica, DollarLaneSelection}. An estimator could also be introduced to learn the nearby traffic volume and velocity to dynamically choose desired speed.

\subsubsection{Vehicle Capabilities}\label{sec:VISSIMCapabilities}
To set realistic capabilities of the powertrains of simulated vehicles, consider the nonlinear maximum acceleration vs. velocity curves as given by the contour regions in Figure \ref{fig:constraint}. We take the maximum acceleration vs velocity curve from a model similar to a Ford Escape and then impose in VISSIM. In this case, a single conventional SUV as modeled from Autonomie is used \citep{autonomie}. We impose only a single powertrain in VISSIM as passenger vehicles rarely exert their maximum capabilities. 

Later, when processing simulation data, a heterogeneous fleet of differing passenger vehicles is considered for fuel economy evaluation - as detailed next.

\subsubsection{Autonomie Energy Evaluation}\label{sec:Autonomie}
To accurately estimate the energy impacts of the proposed control, each vehicle trajectory in the case study was run in Autonomie, a state-of-the-art vehicle energy consumption model first introduced in \citep{autonomie}. Vehicle models in Autonomie are Simulink-based and forward-looking, where virtual pedals are actuated to follow a drive cycle - a speed trajectory as a function of time. Each trajectory is preceded by a model \emph{warm-up}: a speed ramp from 0 to the initial speed of each trajectory, as Autonomie models are designed to start from zero speed. This warm-up portion of the simulation is not considered in the analysis.

Three powertrain configurations were used in post processing: conventional engine-powered vehicle (CV), electric vehicle (EV), and hybrid-electric vehicle (HEV). The HEV is a one-mode power-split hybrid, a configuration similar to the one featured on the Toyota Prius. Each vehicle is of a midsize SUV class, and the component power and mass were sized for each vehicle to meet similar performance requirements, such as 0-60 mph time or ability to climb grades. In addition, the EV was sized to reach a 200 all-electric mile range. Efficiency and power density assumptions are based on Department of Energy technology assumptions for current (2019) vehicles \citep{extensivestudy}. 
\begin{table*}[]
    \centering
    \caption{Vehicle specifications for the Autonomie models considered.}
    \label{tab:autonomie}
    \begin{tabular}{l|c|c|c}
        \hline
         & CV & EV & HEV \\
        \hline
        \hline
        Mass & 1868 kg & 2110 kg & 1938 kg \\
        Aero & $C_DA_f$ = 1.08 m$^2$ & $C_DA_f$ = 1.08 m$^2$ & $C_DA_f$ = 1.08 m$^2$ \\
        Tire & $C_{R}$ = 0.009 & $C_{R}$ = 0.009 & $C_{R}$ = 0.009 \\
        Motor(s) & - & 140 kW, $\eta_{max}$ = 0.98 & 85 kW (M1), 66 kW (M2), $\eta_{max}$ = 0.96 \\
        Engine & 154 kW, $\eta_{max}$ = 0.36 & - & 82 kW, $\eta_{max}$ = 0.41 \\
        \hline
    \end{tabular}
\end{table*}
Table \ref{tab:autonomie} summarizes the specifications for the main vehicles in consideration. Here, $C_D$ is the coefficient of drag, $A_f$ is the frontal drag area, $C_R$ is the coefficient of rolling resistance, and $\eta$ is the efficiency.

\subsubsection{Delay Considerations}
To consider delay due to communication and computation of the systems, first we observe at most {\raise.17ex\hbox{$\scriptstyle\sim$}}10 ms delay from combined optimization building and solving routines. Then, we assume {\raise.17ex\hbox{$\scriptstyle\sim$}}40 ms of delay for DSRC communication algorithms, which are received 86\% of the time and can be sent twice to improve reliability \citep{FengEmpirical}. Then, observe the probability of at least one message to reach the designated vehicle under line-of-sight conditions as $ P(A \cup B) = P(A) + P(B) - P(A \cap B) = 0.98 $. Thus, we consider broadcast communication information to lag one simulation timestep behind - when successfully received.

%%%%%%%%%%%%%%%%%%%%%%%%%%%%%%%%%%%%%%%%%%%
%%% Simulation Results                  %%%
%%%%%%%%%%%%%%%%%%%%%%%%%%%%%%%%%%%%%%%%%%%
\section{Simulation Results} \label{sec:Simulation Results}

To consider a wide range of traffic conditions that can occur throughout a year or even peak to off-peak hours of a given day, discrete scenarios are run by varying the input volume per hour of vehicles, or input flux, in the network. This is done along the interval $[500, 750, ..., 2000]$ veh/hour/lane, which is selected up to the network capacity for the WIE vehicles. For each input flux considered, CAV penetration is then varied along the interval $[0, 10, ..., 100]$\% of the total fleet composition. CAVs are stochastically assigned by Vissim in this case.

%%%%%%%%%%%%%%%%%%%%%%%%%%%%%%%%%%%%%%%%%%%
\subsection{Effects on Traffic Flow} \label{sec:Traffic Flow Results}

\begin{figure}
    \centering
    \hspace{2em}
    \input{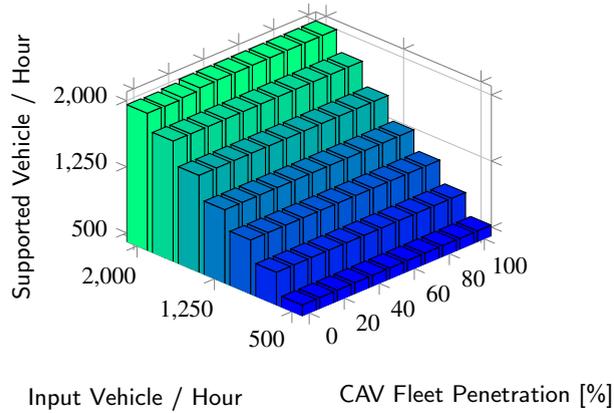}
    \caption{Supported flux of the VISSIM network given an input flux and CAV penetration. Color visualizes the height of each bar.}
    \label{fig:fleet_flux}
\end{figure}

It is important to note that any control algorithm that influences traffic should first examine the road capacity it supports. Reductions in road capacity inhibit travel times and the volume of travellers, effectively lowering the efficiency of the network.

From this, an effective metric in VISSIM to examine road capacity over a discrete grid of simulations is the supported flux of vehicles. This is measured as the total number of vehicles that can be populated in simulation for a given simulation hour. It then follows that if the supported flux of the network is lower than the input flux for a given simulation, then the road capacity has been effectively reduced. This occurs due to jams in the network, which lead to insufficient space at the vehicle source to inject vehicles, and so fewer vehicles are introduced than what is desired.

Figure \ref{fig:fleet_flux} shows the results for the supported flux of vehicles in the network at various input flux and CAV penetrations. It can be observed that in the highest input flux case, the supported vehicle volume per hour is limited in the all-human case by {\raise.17ex\hbox{$\scriptstyle\sim$}}4\%. This trends until 30\% CAV penetration, in which the supported flux in the network improves to match the input flux. Here, with the presence of enough CAVs in the network, the road capacity was improved. 

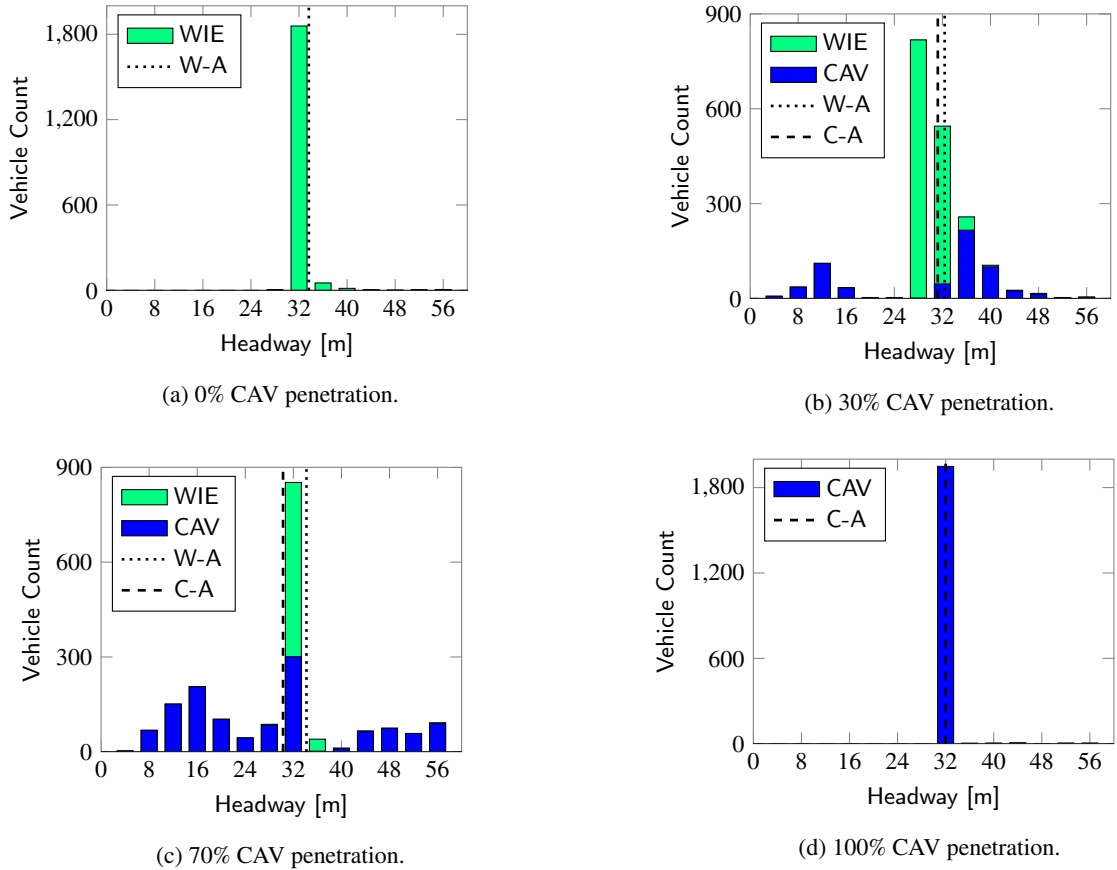
\begin{figure}
    \centering
    
    \begin{subfigure}{0.48\textwidth}
    \hspace{0.5em}
\definecolor{mycolor1}{rgb}{0.00000,1.00000,0.50000}%
\begin{tikzpicture}

\begin{axis}[%
width=4.755cm,
height=3.75cm,
at={(0cm,0cm)},
scale only axis,
% bar shift auto,
xmin=0,
xmax=60,
xtick={ 0,  8, 16, 24, 32, 40, 48, 56},
xlabel style={font=\color{white!15!black}},
xlabel={Headway [m]},
ymin=0,
ymax=2000,
ytick={   0,  600,  1200, 1800},
ylabel style={font=\color{white!15!black}},
ylabel={Vehicle Count},
axis background/.style={fill=white},
legend style={legend pos=north west, legend cell align=left, align=left, draw=black}
]
\addplot[ybar, bar width=6, fill=mycolor1, draw=black, area legend] table[row sep=crcr] {%
0	0\\
4	0\\
8	0\\
12	0\\
16	0\\
20	0\\
24	0\\
28	5\\
32	1858\\
36	52\\
40	14\\
44	4\\
48	2\\
52	4\\
56	6\\
60	0\\
};
\addplot[forget plot, color=white!15!black] table[row sep=crcr] {%
0	0\\
60	0\\
};
\addlegendentry[color=black]{WIE}

\addplot [color=black, dotted, line width=1pt]
  table[row sep=crcr]{%
33.6348782949067	0\\
33.6348782949067	2000\\
};
\addlegendentry[color=black]{W-A}

\end{axis}
\end{tikzpicture}%
    \caption{0\% CAV penetration.}
    \end{subfigure}
    \hfill
    \begin{subfigure}{0.48\textwidth}
    \hspace{1.0em}
\definecolor{mycolor1}{rgb}{0.00000,1.00000,0.50000}%
\begin{tikzpicture}

\begin{axis}[%
width=4.755cm,
height=3.75cm,
at={(0cm,0cm)},
scale only axis,
% bar shift auto,
xmin=0,
xmax=60,
xtick={ 0,  8, 16, 24, 32, 40, 48, 56},
xlabel style={font=\color{white!15!black}},
xlabel={Headway [m]},
ymin=0,
ymax=900,
ytick={  0, 300,600, 900},
ylabel style={font=\color{white!15!black}},
ylabel={Vehicle Count},
axis background/.style={fill=white},
legend style={legend pos=north west, legend cell align=left, align=left, draw=black}
]
\addplot[ybar, bar width=6, fill=mycolor1, draw=black, area legend] table[row sep=crcr] {%
0	0\\
4	7\\
8	36\\
12	111\\
16	34\\
20	3\\
24	3\\
28	818\\
32	545\\
36	258\\
40	105\\
44	26\\
48	16\\
52	3\\
56	5\\
60	0\\
};
\addplot[forget plot, color=white!15!black] table[row sep=crcr] {%
0	0\\
60	0\\
};
\addlegendentry[color=black]{WIE}

\addplot[ybar, bar width=6, fill=blue, draw=black, area legend] table[row sep=crcr] {%
0	0\\
4	7\\
8	36\\
12	111\\
16	34\\
20	3\\
24	3\\
28	1\\
32	46\\
36	216\\
40	100\\
44	24\\
48	12\\
52	3\\
56	2\\
60	0\\
};
\addplot[forget plot, color=white!15!black] table[row sep=crcr] {%
0	0\\
60	0\\
};
\addlegendentry[color=black]{CAV}

\addplot [color=black, dotted, line width=1pt]
  table[row sep=crcr]{%
32.3669091044186	0\\
32.3669091044186	900\\
};
\addlegendentry[color=black]{W-A}

\addplot [color=black, dashed, line width=1pt]
  table[row sep=crcr]{%
31.2311675060012	0\\
31.2311675060012	900\\
};
\addlegendentry[color=black]{C-A}

\end{axis}
\end{tikzpicture}%
    \caption{30\% CAV penetration.}
    \end{subfigure}
    
    \vskip\baselineskip
    
    \begin{subfigure}{0.48\textwidth}
    \hspace{1.0em}
\definecolor{mycolor1}{rgb}{0.00000,1.00000,0.50000}%
\begin{tikzpicture}

\begin{axis}[%
width=4.755cm,
height=3.75cm,
at={(0cm,0cm)},
scale only axis,
% bar shift auto,
xmin=0,
xmax=60,
xtick={ 0,  8, 16, 24, 32, 40, 48, 56},
xlabel style={font=\color{white!15!black}},
xlabel={Headway [m]},
ymin=0,
ymax=900,
ytick={  0, 300, 600, 900},
ylabel style={font=\color{white!15!black}},
ylabel={Vehicle Count},
axis background/.style={fill=white},
legend style={legend pos=north west, legend cell align=left, align=left, draw=black}
]
\addplot[ybar, bar width=6, fill=mycolor1, draw=black, area legend] table[row sep=crcr] {%
0	0\\
4	3\\
8	68\\
12	151\\
16	206\\
20	103\\
24	44\\
28	87\\
32	852\\
36	40\\
40	12\\
44	66\\
48	75\\
52	58\\
56	92\\
60	0\\
};
\addplot[forget plot, color=white!15!black] table[row sep=crcr] {%
0	0\\
60	0\\
};
\addlegendentry[color=black]{WIE}

\addplot[ybar, bar width=6, fill=blue, draw=black, area legend] table[row sep=crcr] {%
0	0\\
4	3\\
8	68\\
12	151\\
16	206\\
20	103\\
24	44\\
28	84\\
32	301\\
36	2\\
40	10\\
44	65\\
48	74\\
52	58\\
56	89\\
60	0\\
};
\addplot[forget plot, color=white!15!black] table[row sep=crcr] {%
0	0\\
60	0\\
};
\addlegendentry[color=black]{CAV}

\addplot [color=black, dotted, line width=1pt]
  table[row sep=crcr]{%
34.1835877167224	0\\
34.1835877167224	900\\
};
\addlegendentry[color=black]{W-A}

\addplot [color=black, dashed, line width=1pt]
  table[row sep=crcr]{%
30.2810376397191	0\\
30.2810376397191	900\\
};
\addlegendentry[color=black]{C-A}

\end{axis}
\end{tikzpicture}%
    \caption{70\% CAV penetration.}
    \end{subfigure}
    \hfill
    \begin{subfigure}{0.48\textwidth}
    \hspace{0.4em}
\begin{tikzpicture}

\begin{axis}[%
width=4.755cm,
height=3.75cm,
at={(0cm,0cm)},
scale only axis,
% bar shift auto,
xmin=0,
xmax=60,
xtick={ 0,  8, 16, 24, 32, 40, 48, 56},
xlabel style={font=\color{white!15!black}},
xlabel={Headway [m]},
ymin=0,
ymax=2000,
ytick={   0,  600,1200, 1800},
ylabel style={font=\color{white!15!black}},
ylabel={Vehicle Count},
axis background/.style={fill=white},
legend style={legend pos=north west, legend cell align=left, align=left, draw=black}
]
\addplot[ybar, bar width=6, fill=blue, draw=black, area legend] table[row sep=crcr] {%
0	0\\
4	0\\
8	0\\
12	0\\
16	0\\
20	0\\
24	0\\
28	0\\
32	1949\\
36	4\\
40	5\\
44	7\\
48	1\\
52	5\\
56	5\\
60	0\\
};
\addplot[forget plot, color=white!15!black] table[row sep=crcr] {%
0	0\\
60	0\\
};
\addlegendentry[color=black]{CAV}

\addplot [color=black, dashed, line width=1pt]
  table[row sep=crcr]{%
32	0\\
32	2000\\
};
\addlegendentry[color=black]{C-A}

\end{axis}
\end{tikzpicture}%
    \caption{100\% CAV penetration.}
    \end{subfigure}
    
    \caption{Vertically stacked histograms of headway of vehicles in the network at various CAV penetrations and 2000 veh/hour. Dashed lines depict the corresponding average to each driver type. It was not possible to populate more than 2000 veh/hour due to the design of VISSIM, so headway increased in CAVs at high penetrations.}
    \label{fig:head_histogram}
\end{figure}

We choose to examine the average space headway of individual drivers at 2000 veh/hour - shown in Figure \ref{fig:head_histogram}. Notably, due to the reduced headway of CAVs with communication benefits, the CAVs are able to achieve better average headway than the WIE drivers with at least 30\% penetration. Thus, a bimodal distribution in the CAVs forms due to anticipative and connected modes - until the 100\% CAV case. It can be observed that the 100\% CAV case creates a unimodal distribution due to increases in the average headway despite all-connected driving. This is due to the combination of the uniform desired speed of the CAVs with the penalty on acceleration in the cost function: they drive under steady conditions once populated in the network and do not attempt to catch their PV despite having safe driving space to do so. In other words, the connected vehicles operate with a commanded headway as demanded by the traffic conditions, and are not operating at the limits of the road capacity of the network. Due to the design of VISSIM in how it populates vehicles in the network, it was not possible to populate more than 2000 vehicles/hour/lane, so it was not possible to study greatly dense networks. 

It should be noted that this result improves on the results observed in other automated driving applications. Our earlier work in \citep{Dollar2018EfficientStrings} chose to assume worst-case braking of the PV for safe prediction. This resulted in a significant loss in road capacity until scenarios with majority CAVs. To address traffic compactness and safety issues explicitly, we introduced the chance constraint approach (Section \ref{sec:Safety}).

%%%%%%%%%%%%%%%%%%%%%%%%%%%%%%%%%%%%%%%%%%%
\subsection{Effects on Fleet Energy Efficiency}\label{sec:FuelEconomy}

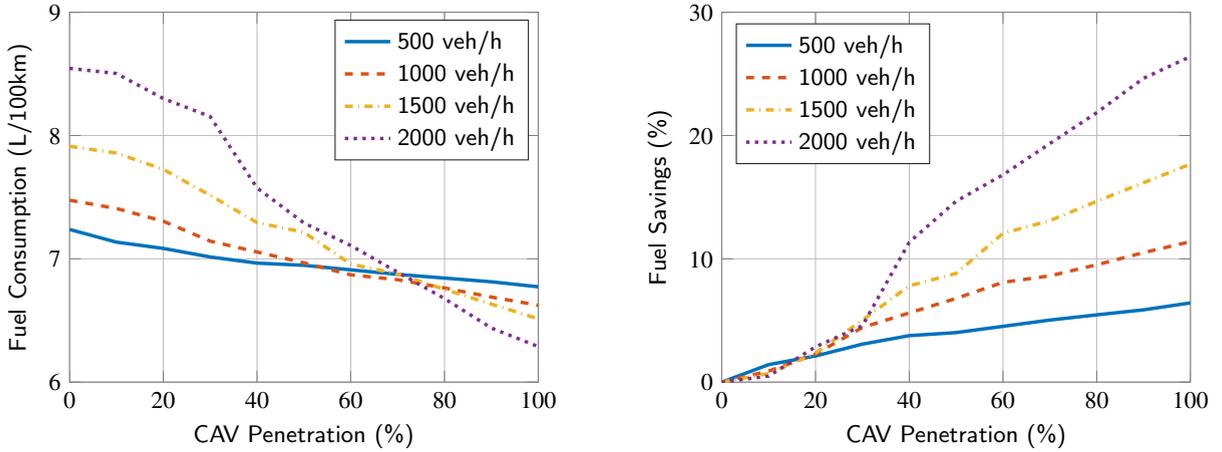
\begin{figure}
    \centering
    
    \begin{subfigure}{0.48\textwidth}
\definecolor{mycolor1}{rgb}{0.00000,0.44700,0.74100}%
\definecolor{mycolor2}{rgb}{0.85000,0.32500,0.09800}%
\definecolor{mycolor3}{rgb}{0.92900,0.69400,0.12500}%
\definecolor{mycolor4}{rgb}{0.49400,0.18400,0.55600}%
\begin{tikzpicture}

\begin{axis}[%
width=6.181cm,
height=4.875cm,
at={(0cm,0cm)},
scale only axis,
xmin=0,
xmax=100,
xlabel style={font=\color{white!15!black}},
xlabel={CAV Penetration (\%)},
ymin=6,
ymax=9,
ylabel style={font=\color{white!15!black}},
ylabel={Fuel Consumption (L/100km)},
axis background/.style={fill=white},
xmajorgrids,
ymajorgrids,
legend style={legend cell align=left, align=left, draw=white!15!black}
]
\addplot [color=mycolor1, line width=1.3pt]
  table[row sep=crcr]{%
0	7.23833083499996\\
10	7.13541494359416\\
20	7.08466223465729\\
30	7.0151833533068\\
40	6.96532598737338\\
50	6.94746642797149\\
70	6.87396158487783\\
90	6.8137928136819\\
100	6.77321020730815\\
};
\addlegendentry[color=black]{500 veh/h}

\addplot [color=mycolor2, dashed, line width=1.3pt]
  table[row sep=crcr]{%
0	7.4753175670468\\
10	7.40846658334044\\
20	7.3050836947773\\
30	7.14380722435232\\
50	6.96759292618501\\
60	6.8702844087157\\
70	6.83092086590699\\
80	6.7637490261878\\
90	6.68989273462525\\
100	6.6238721269839\\
};
\addlegendentry[color=black]{1000 veh/h}

\addplot [color=mycolor3, dashdotted, line width=1.3pt]
  table[row sep=crcr]{%
0	7.91267352618185\\
10	7.85771283193041\\
20	7.7255480756459\\
30	7.51526848414424\\
40	7.29431435917078\\
50	7.21544344958788\\
60	6.95754588888252\\
70	6.87646246468886\\
80	6.75249176096422\\
100	6.51363648938228\\
};
\addlegendentry[color=black]{1500 veh/h}

\addplot [color=mycolor4, dotted, line width=1.3pt]
  table[row sep=crcr]{%
0	8.54486211463302\\
10	8.50315991081911\\
20	8.30008124110039\\
30	8.15436400119931\\
40	7.57494053449156\\
50	7.29099382472008\\
60	7.10775245913098\\
70	6.89167550113575\\
80	6.67744278786429\\
90	6.43881486003971\\
100	6.28851780042632\\
};
\addlegendentry[color=black]{2000 veh/h}

\end{axis}

\begin{axis}[%
width=7.975cm,
height=5.982cm,
at={(-1.037cm,-0.658cm)},
scale only axis,
xmin=0,
xmax=1,
ymin=0,
ymax=1,
axis line style={draw=none},
ticks=none,
axis x line*=bottom,
axis y line*=left,
legend style={legend cell align=left, align=left, draw=white!15!black}
]
\end{axis}
\end{tikzpicture}%
    %\caption{0\% CAV penetration.}
    \end{subfigure}
    \hfill
    \begin{subfigure}{0.48\textwidth}
\definecolor{mycolor1}{rgb}{0.00000,0.44700,0.74100}%
\definecolor{mycolor2}{rgb}{0.85000,0.32500,0.09800}%
\definecolor{mycolor3}{rgb}{0.92900,0.69400,0.12500}%
\definecolor{mycolor4}{rgb}{0.49400,0.18400,0.55600}%
\begin{tikzpicture}
\begin{axis}[%
width=6.181cm,
height=4.875cm,
at={(0cm,0cm)},
scale only axis,
xmin=0,
xmax=100,
xlabel style={font=\color{white!15!black}},
xlabel={CAV Penetration (\%)},
ymin=0,
ymax=30,
ylabel style={font=\color{white!15!black}},
ylabel={Fuel Savings (\%)},
axis background/.style={fill=white},
xmajorgrids,
ymajorgrids,
legend style={at={(0.03,0.97)}, anchor=north west, legend cell align=left, align=left, draw=white!15!black}
]
\addplot [color=mycolor1, line width=1.3pt]
  table[row sep=crcr]{%
0	0\\
10	1.42181800959084\\
20	2.122983928831\\
30	3.08285828293675\\
40	3.77165473435531\\
50	4.01839061599713\\
70	5.0338849995675\\
80	5.45702299337063\\
90	5.86513701840305\\
100	6.42579951503154\\
};
\addlegendentry[color=black]{500 veh/h}

\addplot [color=mycolor2, dashed, line width=1.3pt]
  table[row sep=crcr]{%
0	0\\
10	0.894289548327137\\
20	2.27727946997119\\
30	4.43473256782934\\
50	6.79201433662142\\
60	8.0937452209153\\
70	8.62032542912263\\
80	9.51890718323176\\
90	10.5069092433467\\
100	11.3900905536951\\
};
\addlegendentry[color=black]{1000 veh/h}

\addplot [color=mycolor3, dashdotted, line width=1.3pt]
  table[row sep=crcr]{%
0	0\\
10	0.694590697689932\\
20	2.36488274053988\\
30	5.02238643768951\\
40	7.8147943923758\\
50	8.81156127934419\\
60	12.0708586565457\\
70	13.0955872002595\\
80	14.6623231879687\\
90	16.1786537553731\\
100	17.6809650008984\\
};
\addlegendentry[color=black]{1500 veh/h}

\addplot [color=mycolor4, dotted, line width=1.3pt]
  table[row sep=crcr]{%
0	0\\
10	0.488038346955861\\
20	2.86465562871334\\
30	4.56997559697284\\
40	11.3509330768541\\
50	14.6739440975378\\
60	16.8184066193532\\
70	19.3471420757766\\
80	21.8542944487165\\
90	24.6469425292037\\
100	26.4058598481387\\
};
\addlegendentry[color=black]{2000 veh/h}

\end{axis}

\begin{axis}[%
width=7.975cm,
height=5.982cm,
at={(-1.037cm,-0.658cm)},
scale only axis,
xmin=0,
xmax=1,
ymin=0,
ymax=1,
axis line style={draw=none},
ticks=none,
axis x line*=bottom,
axis y line*=left,
legend style={legend cell align=left, align=left, draw=white!15!black}
]
\end{axis}
\end{tikzpicture}%
    %\caption{30\% CAV penetration.}
    \end{subfigure}

    \caption{CV average fleet fuel consumption and fuel efficiency by veh/hour and CAV penetration.}
    \label{fig:cv_consumption}
\end{figure}

From the traffic flow of the network at all vehicle fluxes and CAV penetrations, fuel economy results for feasible road conditions can be generated using high-fidelity software Autonomie, as detailed in Section \ref{sec:Autonomie}. Each vehicle flux by CAV penetration scenario was simulated with each vehicle type.

Figure \ref{fig:cv_consumption} shows the average energy consumption and average energy saving results with CV powertrains - as a function of penetration and volume. The average energy saving figure uses the energy consumption at 0\% penetration for the same volume as a reference. 

Interestingly, at 1000 veh/hour, the fuel trend is approximately the same as those discussed and found from our earlier work in \citep{Dollar2018EfficientStrings}, which was limited to a string of 8 vehicles, and traffic demand was simulated with a fixed drive cycle in front of the string. However, this result highlights the benefit of the microsimulation approach detailed in this paper: fuel and traffic capacity results at more traffic demands can be studied. 

Higher penetration rates of CAVs bring greater energy savings for the entire fleet. For example, 50\% and 100\% penetration in the high-volume scenario yields 15\% and 27\% energy savings respectively compared to the scenario with no CAVs. Furthermore, the savings are greater at higher volumes for a given penetration, due to greater speed oscillations and more braking events occurring with more vehicles on the road. As a result, the baseline average energy consumption at high volume is greater than at low volumes. As travel times were normalized to each respective all-WIE simulation for each input flux, high volume scenarios travel with lower average velocity than the low volume scenarios, resulting in greater fuel savings in the high volume cases.

\begin{figure}
    \centering
    
    \begin{subfigure}{0.48\textwidth}
\definecolor{mycolor1}{rgb}{0.00000,0.44700,0.74100}%
\definecolor{mycolor2}{rgb}{0.85000,0.32500,0.09800}%
\definecolor{mycolor3}{rgb}{0.92900,0.69400,0.12500}%
\definecolor{mycolor4}{rgb}{0.49400,0.18400,0.55600}%
\begin{tikzpicture}

\begin{axis}[%
width=6.181cm,
height=4.875cm,
at={(0cm,0cm)},
scale only axis,
xmin=0,
xmax=100,
xlabel style={font=\color{white!15!black}},
xlabel={CAV Penetration (\%)},
ymin=-1,
ymax=9,
ylabel style={font=\color{white!15!black}},
ylabel={Electricity Savings (\%)},
axis background/.style={fill=white},
xmajorgrids,
ymajorgrids,
legend style={at={(0.03,0.97)}, anchor=north west, legend cell align=left, align=left, draw=white!15!black}
]
\addplot [color=mycolor1, line width=1.3pt]
  table[row sep=crcr]{%
0	0\\
10	0.506269164152073\\
20	0.488177903466323\\
30	0.852886528754894\\
40	1.08346261547187\\
50	0.762119319626365\\
60	0.869705817880686\\
70	1.25826248780795\\
80	1.29291213130827\\
90	1.42968353333882\\
100	1.49799273460145\\
};
\addlegendentry[color=black]{500 veh/h}

\addplot [color=mycolor2, dashed, line width=1.3pt]
  table[row sep=crcr]{%
0	0\\
10	0.232253165864819\\
20	0.860827112587387\\
30	1.80367966246062\\
40	2.15457411747688\\
50	2.55848676323799\\
60	3.01619588542349\\
70	3.13567786438274\\
80	3.40924908450364\\
90	3.81749908675954\\
100	4.06287112484179\\
};
\addlegendentry[color=black]{1000 veh/h}

\addplot [color=mycolor3, dashdotted, line width=1.3pt]
  table[row sep=crcr]{%
0	0\\
10	0.158934297222729\\
20	0.307162714251234\\
30	1.36204344705276\\
40	2.23607812430906\\
50	2.56274987183164\\
60	3.23013822397148\\
70	3.99572371322147\\
80	4.32902719239173\\
90	4.84650923199426\\
100	5.44884667302296\\
};
\addlegendentry[color=black]{1500 veh/h}

\addplot [color=mycolor4, dotted, line width=1.3pt]
  table[row sep=crcr]{%
0	0\\
10	0.274176785465684\\
20	-0.702024811972848\\
30	-0.970403153609965\\
40	3.47357798401741\\
50	3.22107364315794\\
60	3.166864700673\\
70	4.60373112764189\\
80	5.78010947977737\\
90	7.72107229738268\\
100	8.3760604786017\\
};
\addlegendentry[color=black]{2000 veh/h}

\end{axis}

\begin{axis}[%
width=7.975cm,
height=5.982cm,
at={(-1.037cm,-0.658cm)},
scale only axis,
xmin=0,
xmax=1,
ymin=0,
ymax=1,
axis line style={draw=none},
ticks=none,
axis x line*=bottom,
axis y line*=left,
legend style={legend cell align=left, align=left, draw=white!15!black}
]
\end{axis}
\end{tikzpicture}%
    \caption{EV efficiency.}
    \end{subfigure}
    \hfill
    \begin{subfigure}{0.48\textwidth}
\definecolor{mycolor1}{rgb}{0.00000,0.44700,0.74100}%
\definecolor{mycolor2}{rgb}{0.85000,0.32500,0.09800}%
\definecolor{mycolor3}{rgb}{0.92900,0.69400,0.12500}%
\definecolor{mycolor4}{rgb}{0.49400,0.18400,0.55600}%
\begin{tikzpicture}

\begin{axis}[%
width=6.181cm,
height=4.875cm,
at={(0cm,0cm)},
scale only axis,
xmin=0,
xmax=100,
xlabel style={font=\color{white!15!black}},
xlabel={CAV Penetration (\%)},
ymin=0,
ymax=14,
ylabel style={font=\color{white!15!black}},
ylabel={Fuel Savings (\%)},
axis background/.style={fill=white},
xmajorgrids,
ymajorgrids,
legend style={at={(0.03,0.97)}, anchor=north west, legend cell align=left, align=left, draw=white!15!black}
]
\addplot [color=mycolor1, line width=1.3pt]
  table[row sep=crcr]{%
0	0\\
10	0.606886847925225\\
20	0.718103646233146\\
30	1.10039994271989\\
40	1.45516866026496\\
50	1.30689004460483\\
60	1.57176159262221\\
70	1.95984831049965\\
80	2.14654938918137\\
90	2.36309600088285\\
100	2.59227659805494\\
};
\addlegendentry[color=black]{500 veh/h}

\addplot [color=mycolor2, dashed, line width=1.3pt]
  table[row sep=crcr]{%
0	0\\
10	0.324667590811472\\
20	1.14406858067927\\
30	2.28164567179887\\
40	2.82687212104355\\
50	3.42417287542989\\
60	4.21515070872864\\
70	4.51411790382892\\
80	5.0868396991916\\
90	5.7172649330101\\
100	6.20014123520222\\
};
\addlegendentry[color=black]{1000 veh/h}

\addplot [color=mycolor3, dashdotted, line width=1.3pt]
  table[row sep=crcr]{%
0	0\\
10	0.539869269451259\\
20	1.00353938404993\\
30	2.29528764382003\\
40	3.45437967945432\\
50	3.74610451934575\\
60	5.14794509523855\\
70	5.93197629952081\\
80	6.56132244207942\\
90	7.37543079504522\\
100	8.22201929940834\\
};
\addlegendentry[color=black]{1500 veh/h}

\addplot [color=mycolor4, dotted, line width=1.3pt]
  table[row sep=crcr]{%
0	0\\
10	1.1577865030944\\
20	1.32278809985281\\
30	1.61469255198712\\
40	5.86841420762347\\
50	6.24249928696385\\
60	6.49007957036012\\
70	7.90579700422161\\
80	9.24948704837033\\
90	11.2180780415779\\
100	12.0133185903705\\
};
\addlegendentry[color=black]{2000 veh/h}

\end{axis}

\begin{axis}[%
width=7.975cm,
height=5.982cm,
at={(-1.037cm,-0.658cm)},
scale only axis,
xmin=0,
xmax=1,
ymin=0,
ymax=1,
axis line style={draw=none},
ticks=none,
axis x line*=bottom,
axis y line*=left,
legend style={legend cell align=left, align=left, draw=white!15!black}
]
\end{axis}
\end{tikzpicture}%
    \caption{HEV efficiency.}
    \end{subfigure}
    
    \caption{EV and HEV average fleet energy efficiencies by veh/hour and CAV penetration.}
    \label{fig:ev_consumption}
\end{figure}
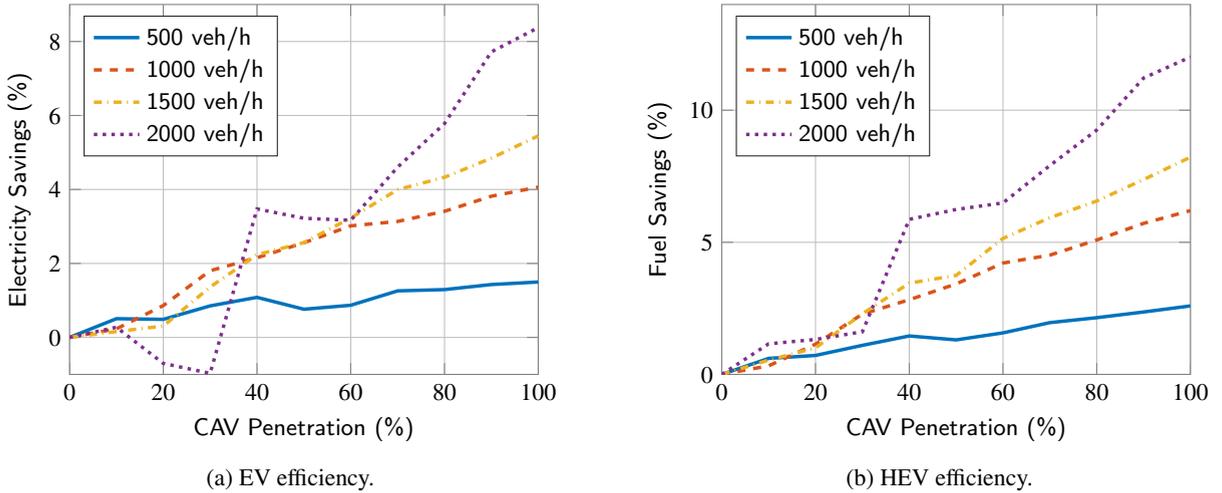

Energy saving trends are similar for the EV and HEV powertrains, as shown in Figure \ref{fig:ev_consumption}, but the magnitude is lower - e.g. the maximum savings for the BEV are 8.5\% and for the HEV are 12\%, whereas maximum savings for the CV are 27\%. The main reason behind this difference is regenerative braking: even in the most congested scenarios at 0\% penetration, no friction braking is necessary for both the BEV and the HEV, and so a large share of the braking energy is recovered by the battery instead, which it can then use for propulsion. In fact, much of the saving potential of hybridization is already achieved at 0\% penetration: the HEV already consumes between 25\% and 35\% less fuel compared to the CV at 0\% penetration rates, and the difference in average consumption between the high and low volume cases at 0\% penetration is low ({\raise.17ex\hbox{$\scriptstyle\sim$}}3\%).

It can be observed EV fleet energy efficiency becomes slightly negative at low CAV penetrations. This is due to slightly worse energy consumptions of human drivers due to improved traffic flow, and thus increased vehicle speeds.

\subsection{Effects on MPC Energy Efficiency}\label{sec:MPCFuelEconomy}
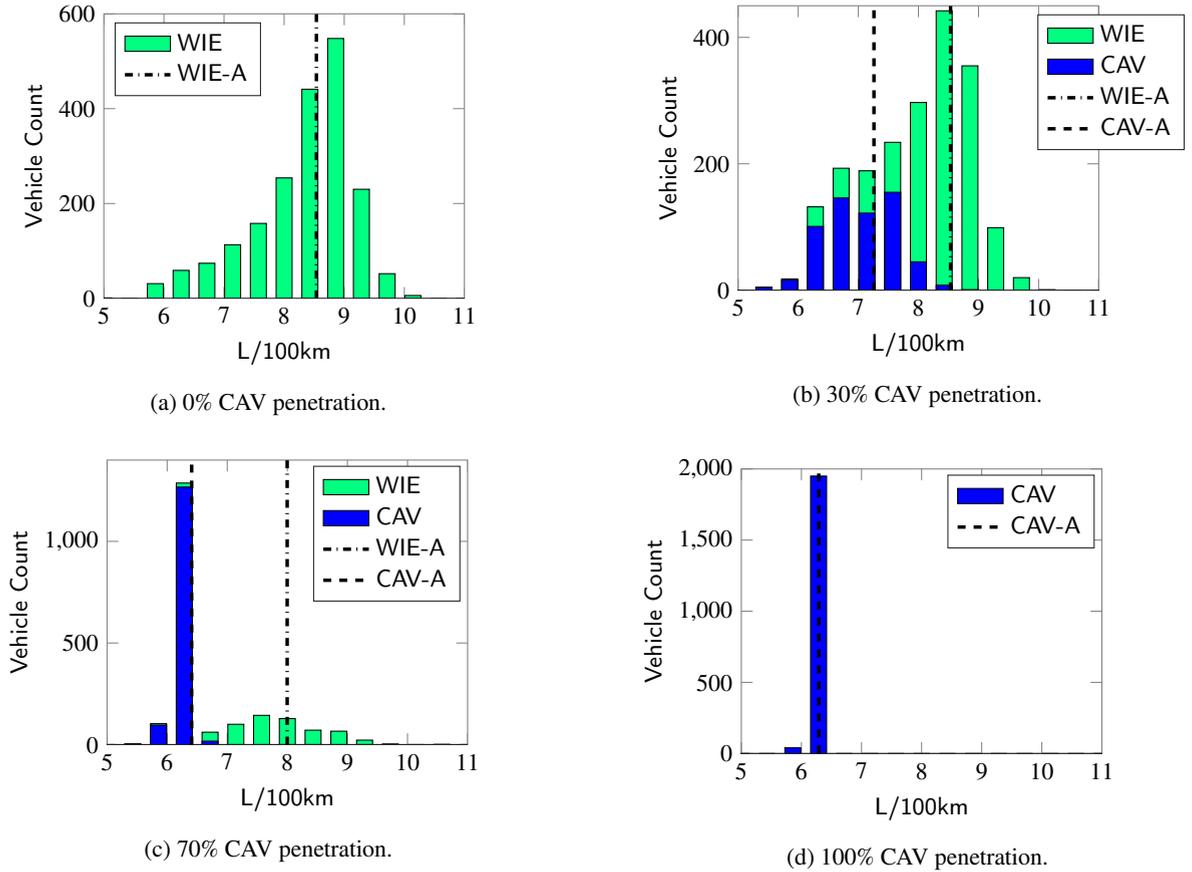
\begin{figure}
    \centering

    \begin{subfigure}{0.48\textwidth}
    \hspace{1.6em}
\definecolor{mycolor1}{rgb}{0.00000,1.00000,0.50000}%
\begin{tikzpicture}

\begin{axis}[%
width=4.755cm,
height=3.75cm,
at={(0cm,0cm)},
scale only axis,
% bar shift auto,
xmin=5,
xmax=11,
xtick={ 5,  6,  7,  8,  9, 10, 11},
xlabel style={font=\color{white!15!black}},
xlabel={L/100km},
ymin=0,
ymax=600,
ylabel style={font=\color{white!15!black}},
ylabel={Vehicle Count},
axis background/.style={fill=white},
legend style={at={(0.03,0.97)}, anchor=north west, legend cell align=left, align=left, draw=white!15!black}
]
\addplot[ybar, bar width=6, fill=mycolor1, draw=black, area legend] table[row sep=crcr] {%
5	1\\
5.42857142857143	0\\
5.85714285714286	31\\
6.28571428571429	59\\
6.71428571428571	74\\
7.14285714285714	113\\
7.57142857142857	158\\
8	254\\
8.42857142857143	441\\
8.85714285714286	548\\
9.28571428571428	230\\
9.71428571428572	52\\
10.1428571428571	6\\
10.5714285714286	0\\
11	0\\
};
\addplot[forget plot, color=white!15!black] table[row sep=crcr] {%
5	0\\
11	0\\
};
\addlegendentry[color=black]{WIE}

\addplot [color=black, dashdotted, line width=1.3pt]
  table[row sep=crcr]{%
8.53737735282175	0\\
8.53737735282175	600\\
};
\addlegendentry[color=black]{WIE-A}

\end{axis}
\end{tikzpicture}%
    \caption{0\% CAV penetration.}
    \end{subfigure}
    \hfill
    \begin{subfigure}{0.48\textwidth}
    \hspace{1.0em}
\definecolor{mycolor1}{rgb}{0.00000,1.00000,0.50000}%
\begin{tikzpicture}

\begin{axis}[%
width=4.755cm,
height=3.75cm,
at={(0cm,0cm)},
scale only axis,
% bar shift auto,
xmin=5,
xmax=11,
xtick={ 5,  6,  7,  8,  9, 10, 11},
xlabel style={font=\color{white!15!black}},
xlabel={L/100km},
ymin=0,
ymax=450,
ylabel style={font=\color{white!15!black}},
ylabel={Vehicle Count},
axis background/.style={fill=white},
legend style={at={(0.83,0.97)}, anchor=north west, legend cell align=left, align=left, draw=white!15!black}
]
\addplot[ybar, bar width=6, fill=mycolor1, draw=black, area legend] table[row sep=crcr] {%
5	0\\
5.42857142857143	5\\
5.85714285714286	18\\
6.28571428571429	132\\
6.71428571428571	193\\
7.14285714285714	189\\
7.57142857142857	234\\
8	297\\
8.42857142857143	442\\
8.85714285714286	355\\
9.28571428571428	99\\
9.71428571428572	20\\
10.1428571428571	1\\
10.5714285714286	0\\
11	0\\
};
\addplot[forget plot, color=white!15!black] table[row sep=crcr] {%
5	0\\
11	0\\
};
\addlegendentry[color=black]{WIE}

\addplot[ybar, bar width=6, fill=blue, draw=black, area legend] table[row sep=crcr] {%
5	0\\
5.42857142857143	5\\
5.85714285714286	16\\
6.28571428571429	101\\
6.71428571428571	146\\
7.14285714285714	122\\
7.57142857142857	155\\
8	45\\
8.42857142857143	8\\
8.85714285714286	1\\
9.28571428571428	0\\
9.71428571428572	0\\
10.1428571428571	0\\
10.5714285714286	0\\
11	0\\
};
\addplot[forget plot, color=white!15!black] table[row sep=crcr] {%
5	0\\
11	0\\
};
\addlegendentry[color=black]{CAV}

\addplot [color=black, dashdotted, line width=1.3pt]
  table[row sep=crcr]{%
8.53494537772241	0\\
8.53494537772241	450\\
};
\addlegendentry[color=black]{WIE-A}

\addplot [color=black, dashed, line width=1.3pt]
  table[row sep=crcr]{%
7.25988012761758	0\\
7.25988012761758	450\\
};
\addlegendentry[color=black]{CAV-A}

\end{axis}
\end{tikzpicture}%
    \caption{30\% CAV penetration.}
    \end{subfigure}
    
    \vskip\baselineskip
    
    \begin{subfigure}{0.48\textwidth}
    \hspace{1.0em}
\definecolor{mycolor1}{rgb}{0.00000,1.00000,0.50000}%
\begin{tikzpicture}

\begin{axis}[%
width=4.755cm,
height=3.75cm,
at={(0cm,0cm)},
scale only axis,
% bar shift auto,
xmin=5,
xmax=11,
xtick={ 5,  6,  7,  8,  9, 10, 11},
xlabel style={font=\color{white!15!black}},
xlabel={L/100km},
ymin=0,
ymax=1400,
ylabel style={font=\color{white!15!black}},
ylabel={Vehicle Count},
axis background/.style={fill=white},
legend style={legend cell align=left, align=left, draw=white!15!black}
]
\addplot[ybar, bar width=6, fill=mycolor1, draw=black, area legend] table[row sep=crcr] {%
5	0\\
5.42857142857143	3\\
5.85714285714286	103\\
6.28571428571429	1287\\
6.71428571428571	61\\
7.14285714285714	100\\
7.57142857142857	144\\
8	128\\
8.42857142857143	71\\
8.85714285714286	66\\
9.28571428571428	22\\
9.71428571428572	3\\
10.1428571428571	0\\
10.5714285714286	1\\
11	0\\
};
\addplot[forget plot, color=white!15!black] table[row sep=crcr] {%
5	0\\
11	0\\
};
\addlegendentry[color=black]{WIE}

\addplot[ybar, bar width=6, fill=blue, draw=black, area legend] table[row sep=crcr] {%
5	0\\
5.42857142857143	3\\
5.85714285714286	96\\
6.28571428571429	1267\\
6.71428571428571	17\\
7.14285714285714	0\\
7.57142857142857	0\\
8	0\\
8.42857142857143	0\\
8.85714285714286	0\\
9.28571428571428	0\\
9.71428571428572	0\\
10.1428571428571	0\\
10.5714285714286	0\\
11	0\\
};
\addplot[forget plot, color=white!15!black] table[row sep=crcr] {%
5	0\\
11	0\\
};
\addlegendentry[color=black]{CAV}

\addplot [color=black, dashdotted, line width=1.3pt]
  table[row sep=crcr]{%
7.99765209573093	0\\
7.99765209573093	1400\\
};
\addlegendentry[color=black]{WIE-A}

\addplot [color=black, dashed, line width=1.3pt]
  table[row sep=crcr]{%
6.4081477491693	0\\
6.4081477491693	1400\\
};
\addlegendentry[color=black]{CAV-A}

\end{axis}
\end{tikzpicture}%
    \caption{70\% CAV penetration.}
    \end{subfigure}
    \hfill
    \begin{subfigure}{0.48\textwidth}
    \hspace{0.4em}
\begin{tikzpicture}

\begin{axis}[%
width=4.755cm,
height=3.75cm,
at={(0cm,0cm)},
scale only axis,
% bar shift auto,
xmin=5,
xmax=11,
xtick={ 5,  6,  7,  8,  9, 10, 11},
xlabel style={font=\color{white!15!black}},
xlabel={L/100km},
ymin=0,
ymax=2000,
ylabel style={font=\color{white!15!black}},
ylabel={Vehicle Count},
axis background/.style={fill=white},
legend style={legend cell align=left, align=left, draw=white!15!black}
]
\addplot[ybar, bar width=6, fill=blue, draw=black, area legend] table[row sep=crcr] {%
5	0\\
5.42857142857143	0\\
5.85714285714286	40\\
6.28571428571429	1950\\
6.71428571428571	0\\
7.14285714285714	0\\
7.57142857142857	0\\
8	0\\
8.42857142857143	0\\
8.85714285714286	0\\
9.28571428571428	0\\
9.71428571428572	0\\
10.1428571428571	0\\
10.5714285714286	0\\
11	0\\
};
\addplot[forget plot, color=white!15!black] table[row sep=crcr] {%
5	0\\
11	0\\
};
\addlegendentry[color=black]{CAV}

\addplot [color=black, dashed, line width=1.3pt]
  table[row sep=crcr]{%
6.28803814197158	0\\
6.28803814197158	2000\\
};
\addlegendentry[color=black]{CAV-A}

\end{axis}
\end{tikzpicture}%
    \caption{100\% CAV penetration.}
    \end{subfigure}
    
    \caption{Histograms of CV fuel economies in the network at various CAV penetrations and 2000 veh/hour. Dashed lines indicate the corresponding average for each driver type.}
    \label{fig:fuel_histogram}
\end{figure}

Then, consider the separation of fuel economies between WIE and CAV, as exemplified with a CV powertrain and 2000 veh/hour in Figure \ref{fig:fuel_histogram}. Here, histograms of the fuel economies of individual vehicles are stacked. There are two general observations to be made here: 1. Fuel economy of individual CAVs at all penetrations is typically much better than that of the WIE vehicles, and 2. Benefits to the fuel economy of WIE vehicles are observed with increasing numbers of CAVs. From this, the WIE vehicles receive secondary benefits to drive with higher fuel efficiency with the presence of CAVs in the fleet. Examining the WIE fuel economies in Figure \ref{fig:fuel_histogram}, at the highest traffic demand and 70\% CAV penetration, the WIE drivers receive up to 10\% average improvement in fuel economy.

Table \ref{tab:fuel} summarizes energy efficiency improvements of the CAVs over WIE drivers at various traffic demands, like those depicted in Figure \ref{fig:fuel_histogram}. This result was consistent across all penetrations of CAVs, $\pm 3\%$ improvement. Notably, the average MPC driver, regardless of hardware, drove more efficiently than the average WIE driver in all scenarios.

\begin{table}[]
    \centering
    \caption{Average energy efficiency improvements of CAVs over WIE drivers with respect to each traffic volume scenario.}
    \label{tab:fuel}
    \begin{tabular}{r|c|c}
        \hline
        Powertrain & Traffic Volume (Veh/H) & Improvement \\
        \hline
        \hline
        CV & Low (500) & 6.6\% \\
        & Med. Low (1000) & 11.3\% \\
        & Med. High (1500) & 18.2\% \\
        & High (2000) & 22.1\% \\
        \hline
        \hline
        EV & Low (500) & 2.8\% \\
        & Med. Low (1000) & 4.1\% \\
        & Med. High (1500) & 5.1\% \\
        & High (2000) & 5.9\% \\
        \hline
        \hline
        HEV & Low (500) & 3.5\% \\
        & Med. Low (1000) & 5.9\% \\
        & Med. High (1500) & 7.5\% \\
        & High (2000) & 8.5\% \\
        \hline
    \end{tabular}
\end{table}

%%%%%%%%%%%%%%%%%%%%%%%%%%%%%%%%%%%%%%%%%%%
\begin{figure}
    \centering
    \ifcompiledensity % Tikz drawings
    \subfloat[0\% CAV penetration.]{
    \input{Images/_CellDensity0.tex}
    }
    \vspace{2pt}
    
    \subfloat[30\% CAV penetration.]{
    \input{Images/_CellDensity30.tex}
    }
    \vspace{2pt}
    
    \subfloat[70\% CAV penetration.]{
    \input{Images/_CellDensity70.tex}
    }
    \vspace{2pt}
    
    \subfloat[100\% CAV penetration.]{
    \input{Images/_CellDensity100.tex}
    }
    \vspace{2pt}
    \else % PNG drawings
    \begin{subfigure}{0.48\textwidth}
    \begin{tikzpicture}
\begin{axis}[%
    xlabel style={font=\color{white!15!black}},
    xlabel={Time [min]},
    xtick={175,1250,2345,3415},
    xticklabels={{4},{21},{39},{56}},
    ylabel style={font=\color{white!15!black}},
    ylabel={Position [km]},
    ytick={2, 7.63, 13.36, 19},
    yticklabels={{0.5},{1.4},{2.2},{3.2}},
    enlargelimits=false, 
    axis on top,
    width=6.5cm,
    colormap={mymap}{[1pt] rgb(0pt)=(0.2422,0.1504,0.6603); rgb(1pt)=(0.25039,0.164995,0.707614); rgb(2pt)=(0.257771,0.181781,0.751138); rgb(3pt)=(0.264729,0.197757,0.795214); rgb(4pt)=(0.270648,0.214676,0.836371); rgb(5pt)=(0.275114,0.234238,0.870986); rgb(6pt)=(0.2783,0.255871,0.899071); rgb(7pt)=(0.280333,0.278233,0.9221); rgb(8pt)=(0.281338,0.300595,0.941376); rgb(9pt)=(0.281014,0.322757,0.957886); rgb(10pt)=(0.279467,0.344671,0.971676); rgb(11pt)=(0.275971,0.366681,0.982905); rgb(12pt)=(0.269914,0.3892,0.9906); rgb(13pt)=(0.260243,0.412329,0.995157); rgb(14pt)=(0.244033,0.435833,0.998833); rgb(15pt)=(0.220643,0.460257,0.997286); rgb(16pt)=(0.196333,0.484719,0.989152); rgb(17pt)=(0.183405,0.507371,0.979795); rgb(18pt)=(0.178643,0.528857,0.968157); rgb(19pt)=(0.176438,0.549905,0.952019); rgb(20pt)=(0.168743,0.570262,0.935871); rgb(21pt)=(0.154,0.5902,0.9218); rgb(22pt)=(0.146029,0.609119,0.907857); rgb(23pt)=(0.138024,0.627629,0.89729); rgb(24pt)=(0.124814,0.645929,0.888343); rgb(25pt)=(0.111252,0.6635,0.876314); rgb(26pt)=(0.0952095,0.679829,0.859781); rgb(27pt)=(0.0688714,0.694771,0.839357); rgb(28pt)=(0.0296667,0.708167,0.816333); rgb(29pt)=(0.00357143,0.720267,0.7917); rgb(30pt)=(0.00665714,0.731214,0.766014); rgb(31pt)=(0.0433286,0.741095,0.73941); rgb(32pt)=(0.0963952,0.75,0.712038); rgb(33pt)=(0.140771,0.7584,0.684157); rgb(34pt)=(0.1717,0.766962,0.655443); rgb(35pt)=(0.193767,0.775767,0.6251); rgb(36pt)=(0.216086,0.7843,0.5923); rgb(37pt)=(0.246957,0.791795,0.556743); rgb(38pt)=(0.290614,0.79729,0.518829); rgb(39pt)=(0.340643,0.8008,0.478857); rgb(40pt)=(0.3909,0.802871,0.435448); rgb(41pt)=(0.445629,0.802419,0.390919); rgb(42pt)=(0.5044,0.7993,0.348); rgb(43pt)=(0.561562,0.794233,0.304481); rgb(44pt)=(0.617395,0.787619,0.261238); rgb(45pt)=(0.671986,0.779271,0.2227); rgb(46pt)=(0.7242,0.769843,0.191029); rgb(47pt)=(0.773833,0.759805,0.16461); rgb(48pt)=(0.820314,0.749814,0.153529); rgb(49pt)=(0.863433,0.7406,0.159633); rgb(50pt)=(0.903543,0.733029,0.177414); rgb(51pt)=(0.939257,0.728786,0.209957); rgb(52pt)=(0.972757,0.729771,0.239443); rgb(53pt)=(0.995648,0.743371,0.237148); rgb(54pt)=(0.996986,0.765857,0.219943); rgb(55pt)=(0.995205,0.789252,0.202762); rgb(56pt)=(0.9892,0.813567,0.188533); rgb(57pt)=(0.978629,0.838629,0.176557); rgb(58pt)=(0.967648,0.8639,0.16429); rgb(59pt)=(0.96101,0.889019,0.153676); rgb(60pt)=(0.959671,0.913457,0.142257); rgb(61pt)=(0.962795,0.937338,0.12651); rgb(62pt)=(0.969114,0.960629,0.106362); rgb(63pt)=(0.9769,0.9839,0.0805)},
    colorbar,
    colorbar style={
        ytick = {0.1, 0.5, 0.9},
        yticklabels = {Low, Med., High},
    }
    ]
\addplot graphics [%
    xmin=155,
    xmax=3446,
    ymin=1,
    ymax=20] {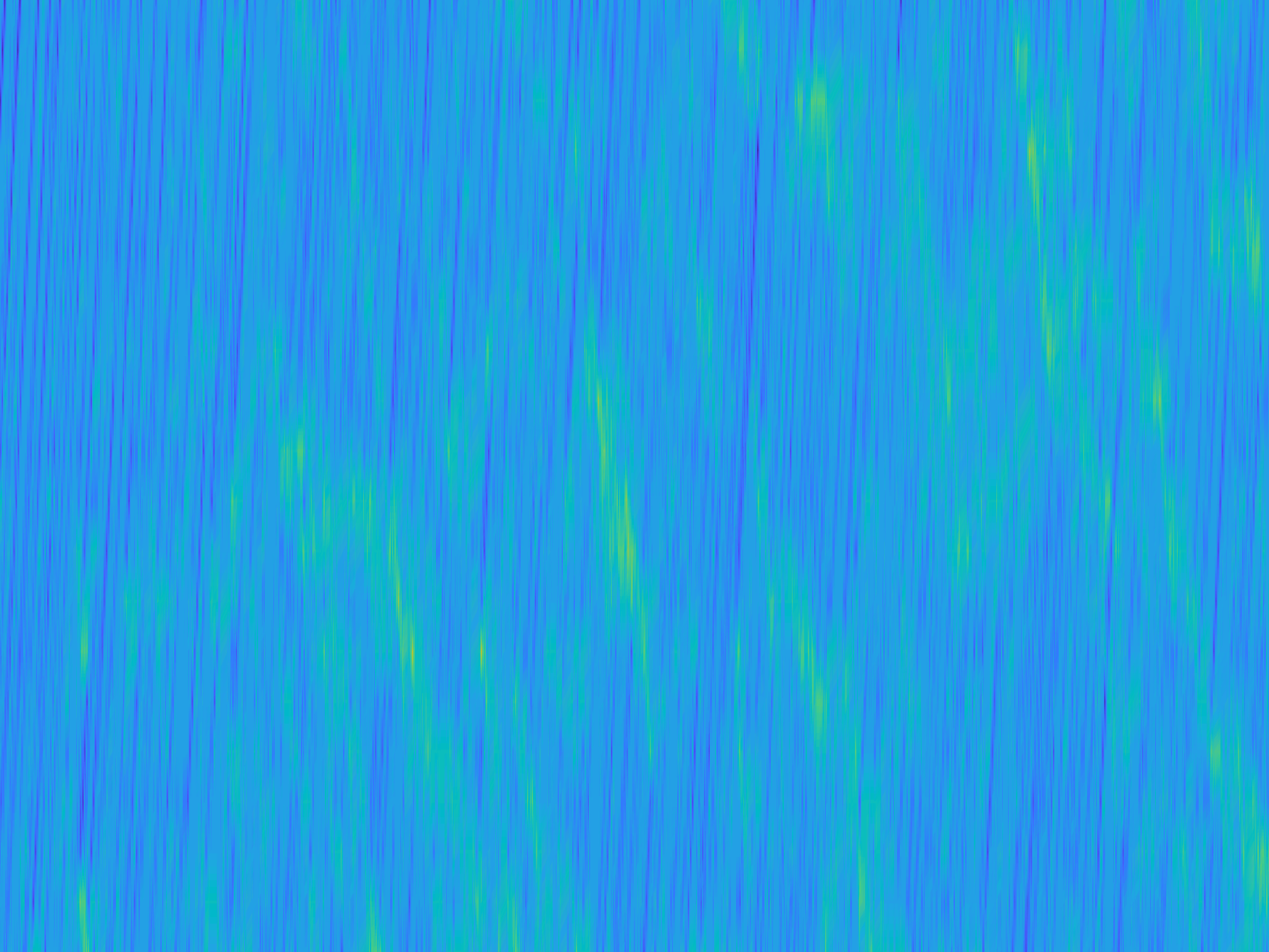};
\end{axis}
\end{tikzpicture}
    \caption{0\% CAV penetration.}
    \end{subfigure}
    \hfill
    \begin{subfigure}{0.48\textwidth}
    \begin{tikzpicture}
\begin{axis}[%
    xlabel style={font=\color{white!15!black}},
    xlabel={Time [min]},
    xtick={175,1250,2345,3415},
    xticklabels={{4},{21},{39},{56}},
    ylabel style={font=\color{white!15!black}},
    ylabel={Position [km]},
    ytick={2, 7.63, 13.36, 19},
    yticklabels={{0.5},{1.4},{2.2},{3.2}},
    enlargelimits=false, 
    axis on top,
    width=6.5cm,
    colormap={mymap}{[1pt] rgb(0pt)=(0.2422,0.1504,0.6603); rgb(1pt)=(0.25039,0.164995,0.707614); rgb(2pt)=(0.257771,0.181781,0.751138); rgb(3pt)=(0.264729,0.197757,0.795214); rgb(4pt)=(0.270648,0.214676,0.836371); rgb(5pt)=(0.275114,0.234238,0.870986); rgb(6pt)=(0.2783,0.255871,0.899071); rgb(7pt)=(0.280333,0.278233,0.9221); rgb(8pt)=(0.281338,0.300595,0.941376); rgb(9pt)=(0.281014,0.322757,0.957886); rgb(10pt)=(0.279467,0.344671,0.971676); rgb(11pt)=(0.275971,0.366681,0.982905); rgb(12pt)=(0.269914,0.3892,0.9906); rgb(13pt)=(0.260243,0.412329,0.995157); rgb(14pt)=(0.244033,0.435833,0.998833); rgb(15pt)=(0.220643,0.460257,0.997286); rgb(16pt)=(0.196333,0.484719,0.989152); rgb(17pt)=(0.183405,0.507371,0.979795); rgb(18pt)=(0.178643,0.528857,0.968157); rgb(19pt)=(0.176438,0.549905,0.952019); rgb(20pt)=(0.168743,0.570262,0.935871); rgb(21pt)=(0.154,0.5902,0.9218); rgb(22pt)=(0.146029,0.609119,0.907857); rgb(23pt)=(0.138024,0.627629,0.89729); rgb(24pt)=(0.124814,0.645929,0.888343); rgb(25pt)=(0.111252,0.6635,0.876314); rgb(26pt)=(0.0952095,0.679829,0.859781); rgb(27pt)=(0.0688714,0.694771,0.839357); rgb(28pt)=(0.0296667,0.708167,0.816333); rgb(29pt)=(0.00357143,0.720267,0.7917); rgb(30pt)=(0.00665714,0.731214,0.766014); rgb(31pt)=(0.0433286,0.741095,0.73941); rgb(32pt)=(0.0963952,0.75,0.712038); rgb(33pt)=(0.140771,0.7584,0.684157); rgb(34pt)=(0.1717,0.766962,0.655443); rgb(35pt)=(0.193767,0.775767,0.6251); rgb(36pt)=(0.216086,0.7843,0.5923); rgb(37pt)=(0.246957,0.791795,0.556743); rgb(38pt)=(0.290614,0.79729,0.518829); rgb(39pt)=(0.340643,0.8008,0.478857); rgb(40pt)=(0.3909,0.802871,0.435448); rgb(41pt)=(0.445629,0.802419,0.390919); rgb(42pt)=(0.5044,0.7993,0.348); rgb(43pt)=(0.561562,0.794233,0.304481); rgb(44pt)=(0.617395,0.787619,0.261238); rgb(45pt)=(0.671986,0.779271,0.2227); rgb(46pt)=(0.7242,0.769843,0.191029); rgb(47pt)=(0.773833,0.759805,0.16461); rgb(48pt)=(0.820314,0.749814,0.153529); rgb(49pt)=(0.863433,0.7406,0.159633); rgb(50pt)=(0.903543,0.733029,0.177414); rgb(51pt)=(0.939257,0.728786,0.209957); rgb(52pt)=(0.972757,0.729771,0.239443); rgb(53pt)=(0.995648,0.743371,0.237148); rgb(54pt)=(0.996986,0.765857,0.219943); rgb(55pt)=(0.995205,0.789252,0.202762); rgb(56pt)=(0.9892,0.813567,0.188533); rgb(57pt)=(0.978629,0.838629,0.176557); rgb(58pt)=(0.967648,0.8639,0.16429); rgb(59pt)=(0.96101,0.889019,0.153676); rgb(60pt)=(0.959671,0.913457,0.142257); rgb(61pt)=(0.962795,0.937338,0.12651); rgb(62pt)=(0.969114,0.960629,0.106362); rgb(63pt)=(0.9769,0.9839,0.0805)},
    colorbar,
    colorbar style={
        ytick = {0.1, 0.5, 0.9},
        yticklabels = {Low, Med., High},
    }
    ]
\addplot graphics [%
    xmin=155,
    xmax=3446,
    ymin=1,
    ymax=20] {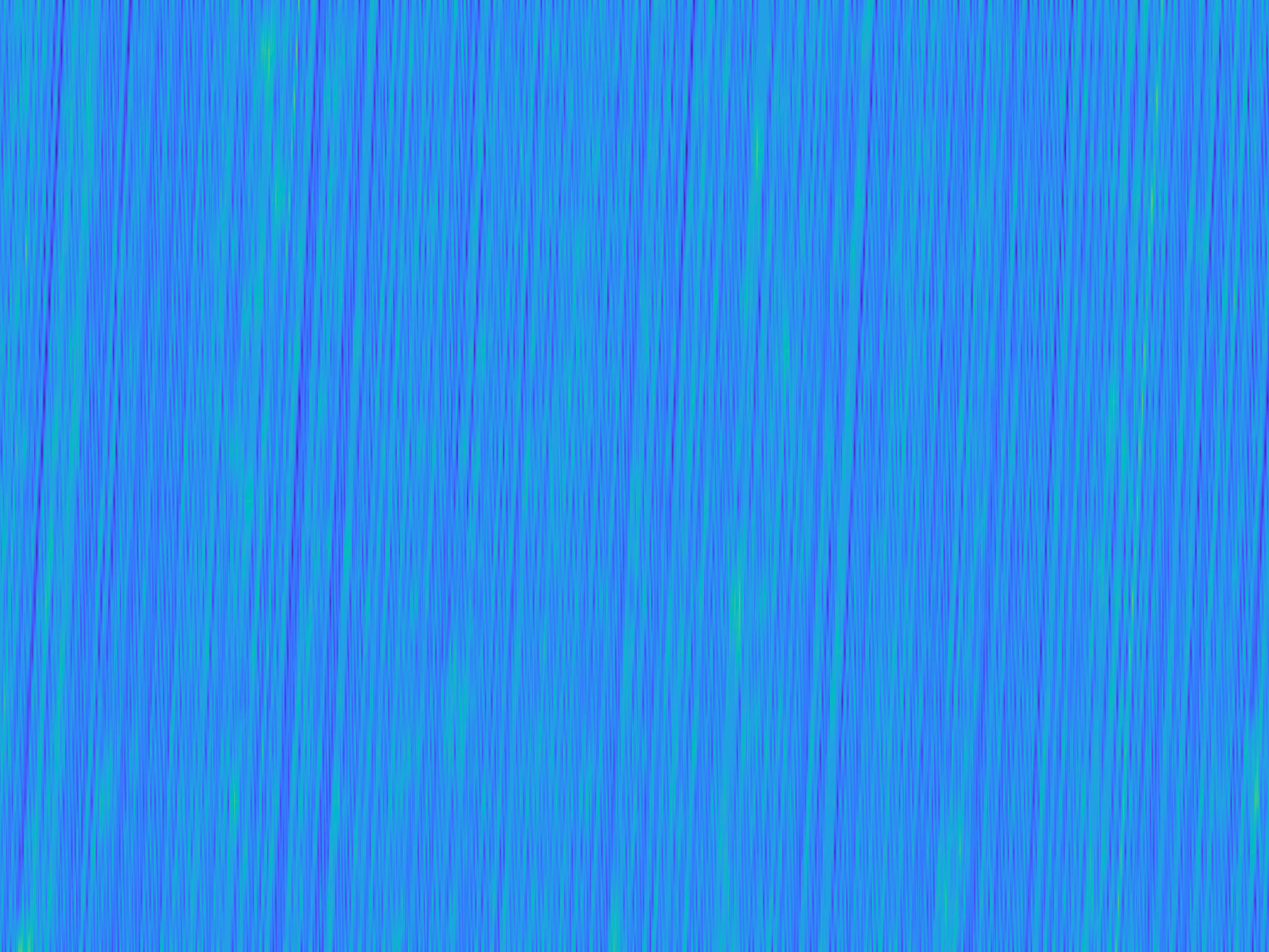};
\end{axis}
\end{tikzpicture}
    \caption{30\% CAV penetration.}
    \end{subfigure}
    
    \vskip\baselineskip
    
    \begin{subfigure}{0.48\textwidth}
    \begin{tikzpicture}
\begin{axis}[%
    xlabel style={font=\color{white!15!black}},
    xlabel={Time [min]},
    xtick={175,1250,2345,3415},
    xticklabels={{4},{21},{39},{56}},
    ylabel style={font=\color{white!15!black}},
    ylabel={Position [km]},
    ytick={2, 7.63, 13.36, 19},
    yticklabels={{0.5},{1.4},{2.2},{3.2}},
    enlargelimits=false, 
    axis on top,
    width=6.5cm,
    colormap={mymap}{[1pt] rgb(0pt)=(0.2422,0.1504,0.6603); rgb(1pt)=(0.25039,0.164995,0.707614); rgb(2pt)=(0.257771,0.181781,0.751138); rgb(3pt)=(0.264729,0.197757,0.795214); rgb(4pt)=(0.270648,0.214676,0.836371); rgb(5pt)=(0.275114,0.234238,0.870986); rgb(6pt)=(0.2783,0.255871,0.899071); rgb(7pt)=(0.280333,0.278233,0.9221); rgb(8pt)=(0.281338,0.300595,0.941376); rgb(9pt)=(0.281014,0.322757,0.957886); rgb(10pt)=(0.279467,0.344671,0.971676); rgb(11pt)=(0.275971,0.366681,0.982905); rgb(12pt)=(0.269914,0.3892,0.9906); rgb(13pt)=(0.260243,0.412329,0.995157); rgb(14pt)=(0.244033,0.435833,0.998833); rgb(15pt)=(0.220643,0.460257,0.997286); rgb(16pt)=(0.196333,0.484719,0.989152); rgb(17pt)=(0.183405,0.507371,0.979795); rgb(18pt)=(0.178643,0.528857,0.968157); rgb(19pt)=(0.176438,0.549905,0.952019); rgb(20pt)=(0.168743,0.570262,0.935871); rgb(21pt)=(0.154,0.5902,0.9218); rgb(22pt)=(0.146029,0.609119,0.907857); rgb(23pt)=(0.138024,0.627629,0.89729); rgb(24pt)=(0.124814,0.645929,0.888343); rgb(25pt)=(0.111252,0.6635,0.876314); rgb(26pt)=(0.0952095,0.679829,0.859781); rgb(27pt)=(0.0688714,0.694771,0.839357); rgb(28pt)=(0.0296667,0.708167,0.816333); rgb(29pt)=(0.00357143,0.720267,0.7917); rgb(30pt)=(0.00665714,0.731214,0.766014); rgb(31pt)=(0.0433286,0.741095,0.73941); rgb(32pt)=(0.0963952,0.75,0.712038); rgb(33pt)=(0.140771,0.7584,0.684157); rgb(34pt)=(0.1717,0.766962,0.655443); rgb(35pt)=(0.193767,0.775767,0.6251); rgb(36pt)=(0.216086,0.7843,0.5923); rgb(37pt)=(0.246957,0.791795,0.556743); rgb(38pt)=(0.290614,0.79729,0.518829); rgb(39pt)=(0.340643,0.8008,0.478857); rgb(40pt)=(0.3909,0.802871,0.435448); rgb(41pt)=(0.445629,0.802419,0.390919); rgb(42pt)=(0.5044,0.7993,0.348); rgb(43pt)=(0.561562,0.794233,0.304481); rgb(44pt)=(0.617395,0.787619,0.261238); rgb(45pt)=(0.671986,0.779271,0.2227); rgb(46pt)=(0.7242,0.769843,0.191029); rgb(47pt)=(0.773833,0.759805,0.16461); rgb(48pt)=(0.820314,0.749814,0.153529); rgb(49pt)=(0.863433,0.7406,0.159633); rgb(50pt)=(0.903543,0.733029,0.177414); rgb(51pt)=(0.939257,0.728786,0.209957); rgb(52pt)=(0.972757,0.729771,0.239443); rgb(53pt)=(0.995648,0.743371,0.237148); rgb(54pt)=(0.996986,0.765857,0.219943); rgb(55pt)=(0.995205,0.789252,0.202762); rgb(56pt)=(0.9892,0.813567,0.188533); rgb(57pt)=(0.978629,0.838629,0.176557); rgb(58pt)=(0.967648,0.8639,0.16429); rgb(59pt)=(0.96101,0.889019,0.153676); rgb(60pt)=(0.959671,0.913457,0.142257); rgb(61pt)=(0.962795,0.937338,0.12651); rgb(62pt)=(0.969114,0.960629,0.106362); rgb(63pt)=(0.9769,0.9839,0.0805)},
    colorbar,
    colorbar style={
        ytick = {0.1, 0.5, 0.9},
        yticklabels = {Low, Med., High},
    }
    ]
\addplot graphics [%
    xmin=155,
    xmax=3446,
    ymin=1,
    ymax=20] {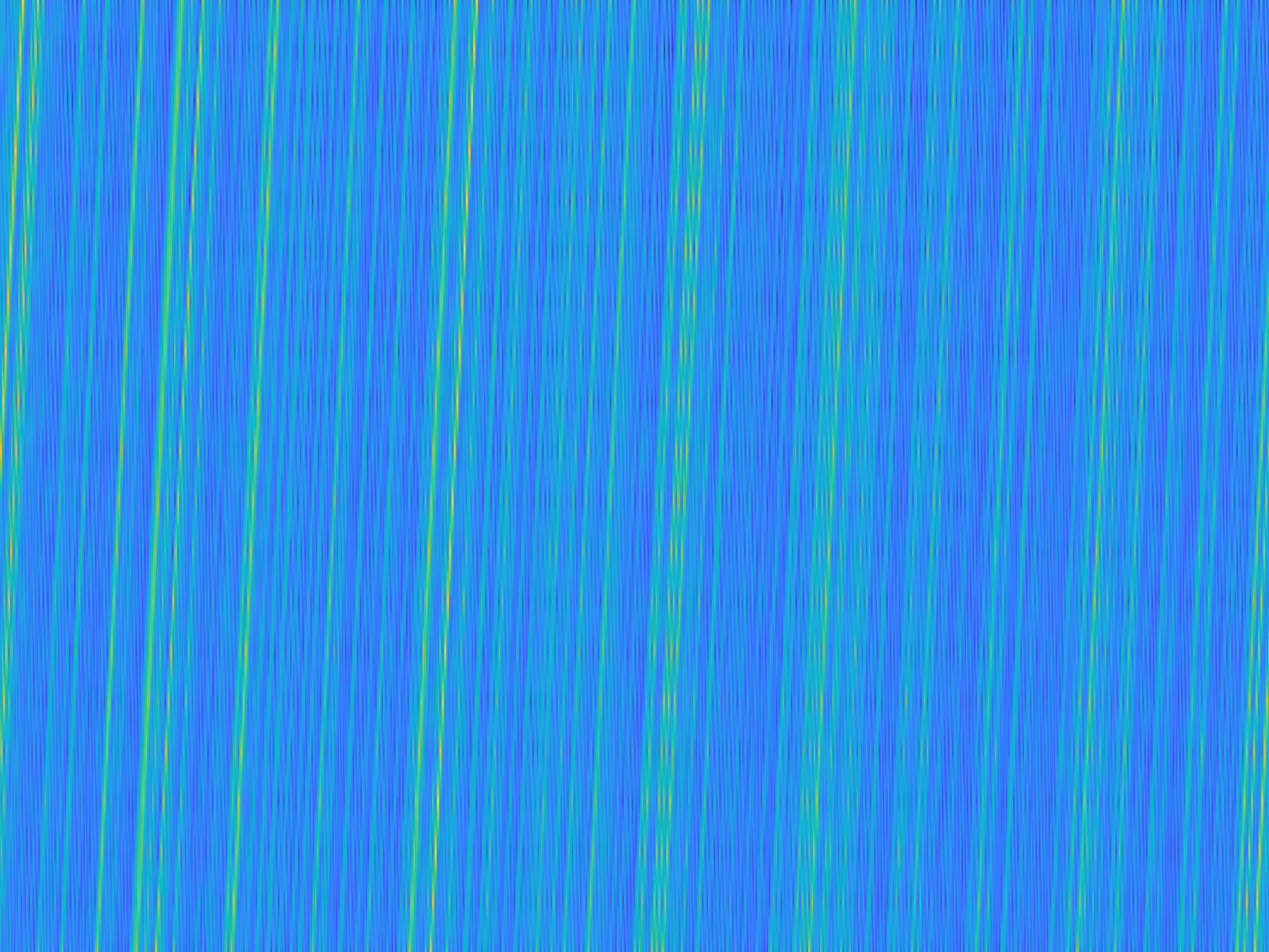};
\end{axis}
\end{tikzpicture}
    \caption{70\% CAV penetration.}
    \end{subfigure}
    \hfill
    \begin{subfigure}{0.48\textwidth}
    \begin{tikzpicture}
\begin{axis}[%
    xlabel style={font=\color{white!15!black}},
    xlabel={Time [min]},
    xtick={175,1250,2345,3415},
    xticklabels={{4},{21},{39},{56}},
    ylabel style={font=\color{white!15!black}},
    ylabel={Position [km]},
    ytick={2, 7.63, 13.36, 19},
    yticklabels={{0.5},{1.4},{2.2},{3.2}},
    enlargelimits=false, 
    axis on top,
    width=6.5cm,
    colormap={mymap}{[1pt] rgb(0pt)=(0.2422,0.1504,0.6603); rgb(1pt)=(0.25039,0.164995,0.707614); rgb(2pt)=(0.257771,0.181781,0.751138); rgb(3pt)=(0.264729,0.197757,0.795214); rgb(4pt)=(0.270648,0.214676,0.836371); rgb(5pt)=(0.275114,0.234238,0.870986); rgb(6pt)=(0.2783,0.255871,0.899071); rgb(7pt)=(0.280333,0.278233,0.9221); rgb(8pt)=(0.281338,0.300595,0.941376); rgb(9pt)=(0.281014,0.322757,0.957886); rgb(10pt)=(0.279467,0.344671,0.971676); rgb(11pt)=(0.275971,0.366681,0.982905); rgb(12pt)=(0.269914,0.3892,0.9906); rgb(13pt)=(0.260243,0.412329,0.995157); rgb(14pt)=(0.244033,0.435833,0.998833); rgb(15pt)=(0.220643,0.460257,0.997286); rgb(16pt)=(0.196333,0.484719,0.989152); rgb(17pt)=(0.183405,0.507371,0.979795); rgb(18pt)=(0.178643,0.528857,0.968157); rgb(19pt)=(0.176438,0.549905,0.952019); rgb(20pt)=(0.168743,0.570262,0.935871); rgb(21pt)=(0.154,0.5902,0.9218); rgb(22pt)=(0.146029,0.609119,0.907857); rgb(23pt)=(0.138024,0.627629,0.89729); rgb(24pt)=(0.124814,0.645929,0.888343); rgb(25pt)=(0.111252,0.6635,0.876314); rgb(26pt)=(0.0952095,0.679829,0.859781); rgb(27pt)=(0.0688714,0.694771,0.839357); rgb(28pt)=(0.0296667,0.708167,0.816333); rgb(29pt)=(0.00357143,0.720267,0.7917); rgb(30pt)=(0.00665714,0.731214,0.766014); rgb(31pt)=(0.0433286,0.741095,0.73941); rgb(32pt)=(0.0963952,0.75,0.712038); rgb(33pt)=(0.140771,0.7584,0.684157); rgb(34pt)=(0.1717,0.766962,0.655443); rgb(35pt)=(0.193767,0.775767,0.6251); rgb(36pt)=(0.216086,0.7843,0.5923); rgb(37pt)=(0.246957,0.791795,0.556743); rgb(38pt)=(0.290614,0.79729,0.518829); rgb(39pt)=(0.340643,0.8008,0.478857); rgb(40pt)=(0.3909,0.802871,0.435448); rgb(41pt)=(0.445629,0.802419,0.390919); rgb(42pt)=(0.5044,0.7993,0.348); rgb(43pt)=(0.561562,0.794233,0.304481); rgb(44pt)=(0.617395,0.787619,0.261238); rgb(45pt)=(0.671986,0.779271,0.2227); rgb(46pt)=(0.7242,0.769843,0.191029); rgb(47pt)=(0.773833,0.759805,0.16461); rgb(48pt)=(0.820314,0.749814,0.153529); rgb(49pt)=(0.863433,0.7406,0.159633); rgb(50pt)=(0.903543,0.733029,0.177414); rgb(51pt)=(0.939257,0.728786,0.209957); rgb(52pt)=(0.972757,0.729771,0.239443); rgb(53pt)=(0.995648,0.743371,0.237148); rgb(54pt)=(0.996986,0.765857,0.219943); rgb(55pt)=(0.995205,0.789252,0.202762); rgb(56pt)=(0.9892,0.813567,0.188533); rgb(57pt)=(0.978629,0.838629,0.176557); rgb(58pt)=(0.967648,0.8639,0.16429); rgb(59pt)=(0.96101,0.889019,0.153676); rgb(60pt)=(0.959671,0.913457,0.142257); rgb(61pt)=(0.962795,0.937338,0.12651); rgb(62pt)=(0.969114,0.960629,0.106362); rgb(63pt)=(0.9769,0.9839,0.0805)},
    colorbar,
    colorbar style={
        ytick = {0.1, 0.5, 0.9},
        yticklabels = {Low, Med., High},
    }
    ]
\addplot graphics [%
    xmin=155,
    xmax=3446,
    ymin=1,
    ymax=20] {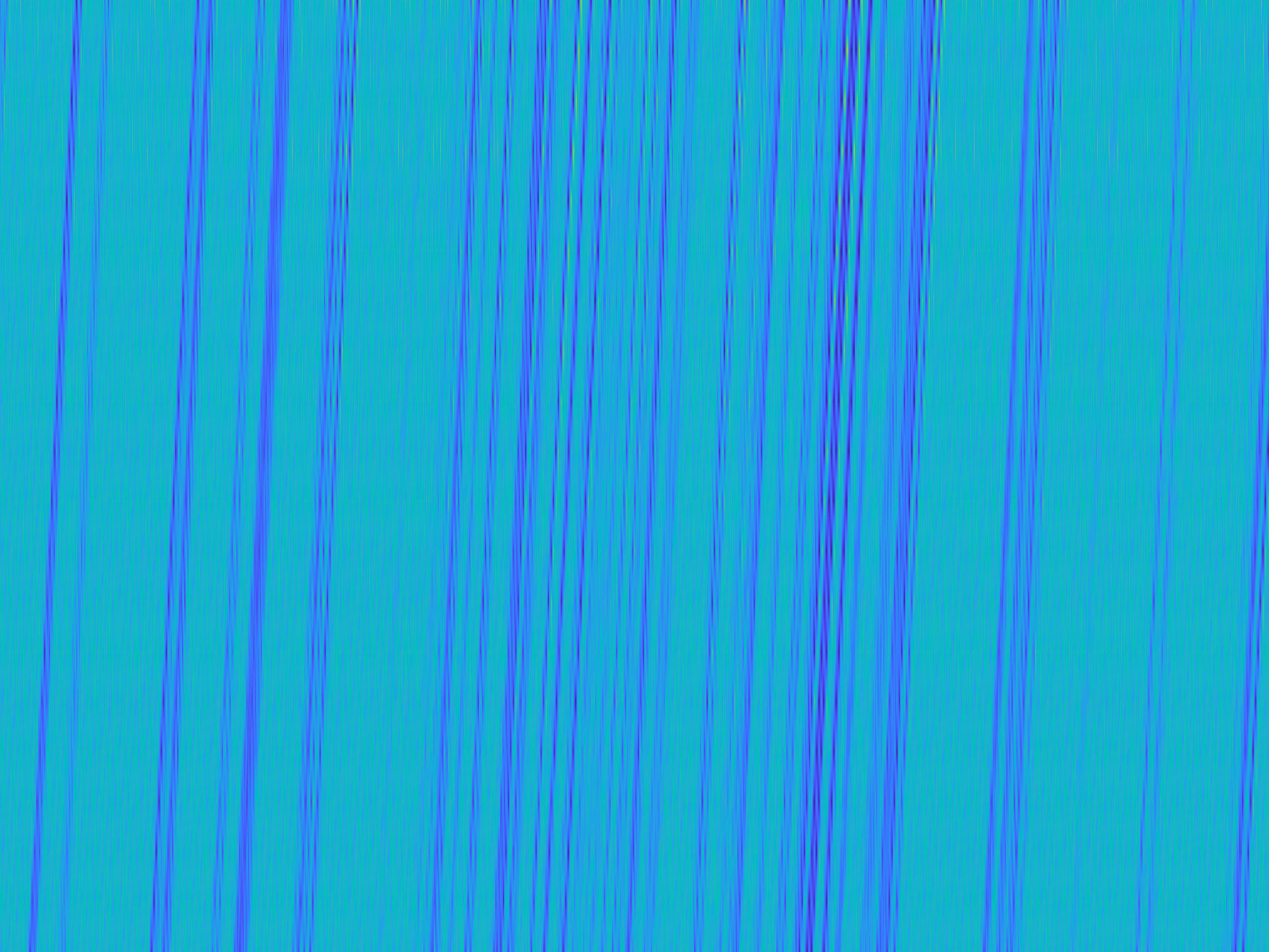};
\end{axis}
\end{tikzpicture}
    \caption{100\% CAV penetration.}
    \end{subfigure}
    \fi
    \caption{Cell density plots at various CAV penetrations and 2000 veh/hour. Shockwaves propagate backwards at 0\%.}
    \label{fig:cell_density}
\end{figure}

\subsection{Effects on Driving Behavior}
Clearly, energy benefits observed are significant and corroborate with results offered by other studies such as \citep{LiuFreeway}. To further study and explain some phenomena as to why they occurred, we can first examine traffic smoothing effects through dissipation of shockwaves.

Figure \ref{fig:cell_density} depicts a series of cell density plots at 2000 veh/hour. Here, the density of vehicles measured within a discrete set of cells of the road network are plotted as colored regions over position and time. At 0\% CAV penetration, an observed behavior are shockwaves in the network, shown as high density regions propagating backwards through the network over time. Stabilizing effects are then observed with the introduction of CAVs. Here, traffic is smoothed and the shockwaves are dissipated - as seen with 30\% CAVs. Further, penetrations of 70\% and 100\% CAVs result in uniform platoons of vehicles moving forward through the network over time.

The vehicle control proposed here promotes secondary benefits of smoothing traffic. From this, reducing shockwave effects in the network created less stop-and-go behavior, and so fewer acceleration and braking events are experienced by vehicles. Therefore, energy consumption is reduced for involved vehicles.

\begin{figure}
    \centering
    
    \begin{subfigure}{0.48\textwidth}
    \hspace{1.0em}
    \begin{tikzpicture}
\begin{axis}[%
    xlabel style={font=\color{white!15!black}},
    xlabel={Time [min]},
    xtick={175,1250,2345,3415},
    xticklabels={{4},{21},{39},{56}},
    ylabel style={font=\color{white!15!black}},
    ylabel={Velocity [m/s]},
    enlargelimits=false, 
    axis on top,  
    width=6.5cm,
    legend style={
        draw=white!15!black,
        legend cell align=left,
        legend pos=south east}
    ]
    \addlegendimage{blue,mark=*}
    \addlegendentry[color=black]{Velocity Sample};
    \addlegendimage{red}
    \addlegendentry[color=black]{Average};
    ]
\addplot graphics [%
    xmin=107,
    xmax=3600,
    ymin=0,
    ymax=28] {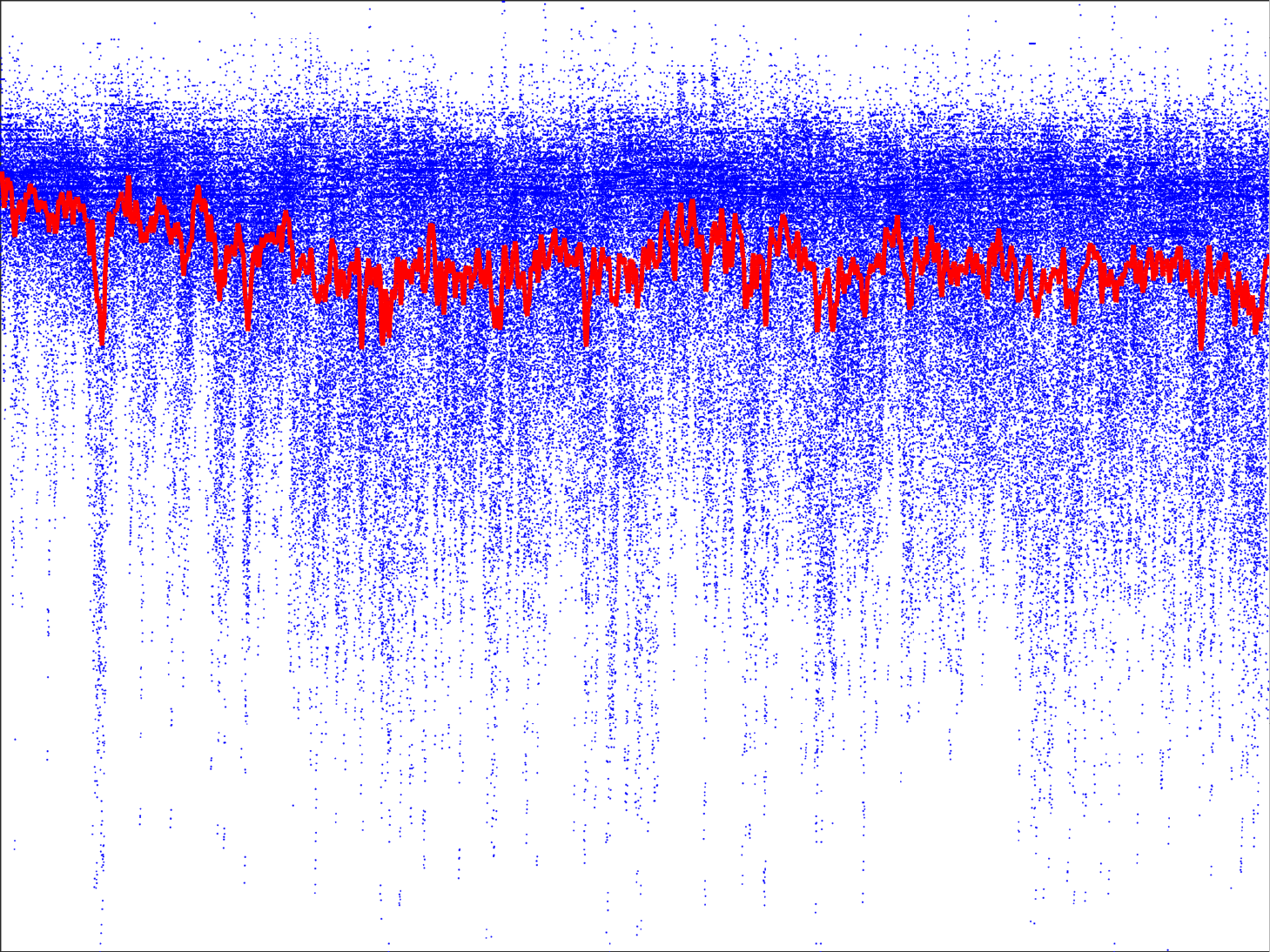};
\end{axis}
\end{tikzpicture}
    \caption{0\% CAV penetration.}
    \end{subfigure}
    \hfill
    \begin{subfigure}{0.48\textwidth}
    \hspace{1.0em}
    \begin{tikzpicture}
\begin{axis}[%
    xlabel style={font=\color{white!15!black}},
    xlabel={Time [min]},
    xtick={175,1250,2345,3415},
    xticklabels={{4},{21},{39},{56}},
    ylabel style={font=\color{white!15!black}},
    ylabel={Velocity [m/s]},
    enlargelimits=false, 
    axis on top,  
    width=6.5cm,
    legend style={
        draw=white!15!black,
        legend cell align=left,
        legend pos=south east}
    ]
    \addlegendimage{blue,mark=*}
    \addlegendentry[color=black]{Velocity Sample};
    \addlegendimage{red}
    \addlegendentry[color=black]{Average};
    ]
\addplot graphics [%
    xmin=107,
    xmax=3600,
    ymin=0,
    ymax=28] {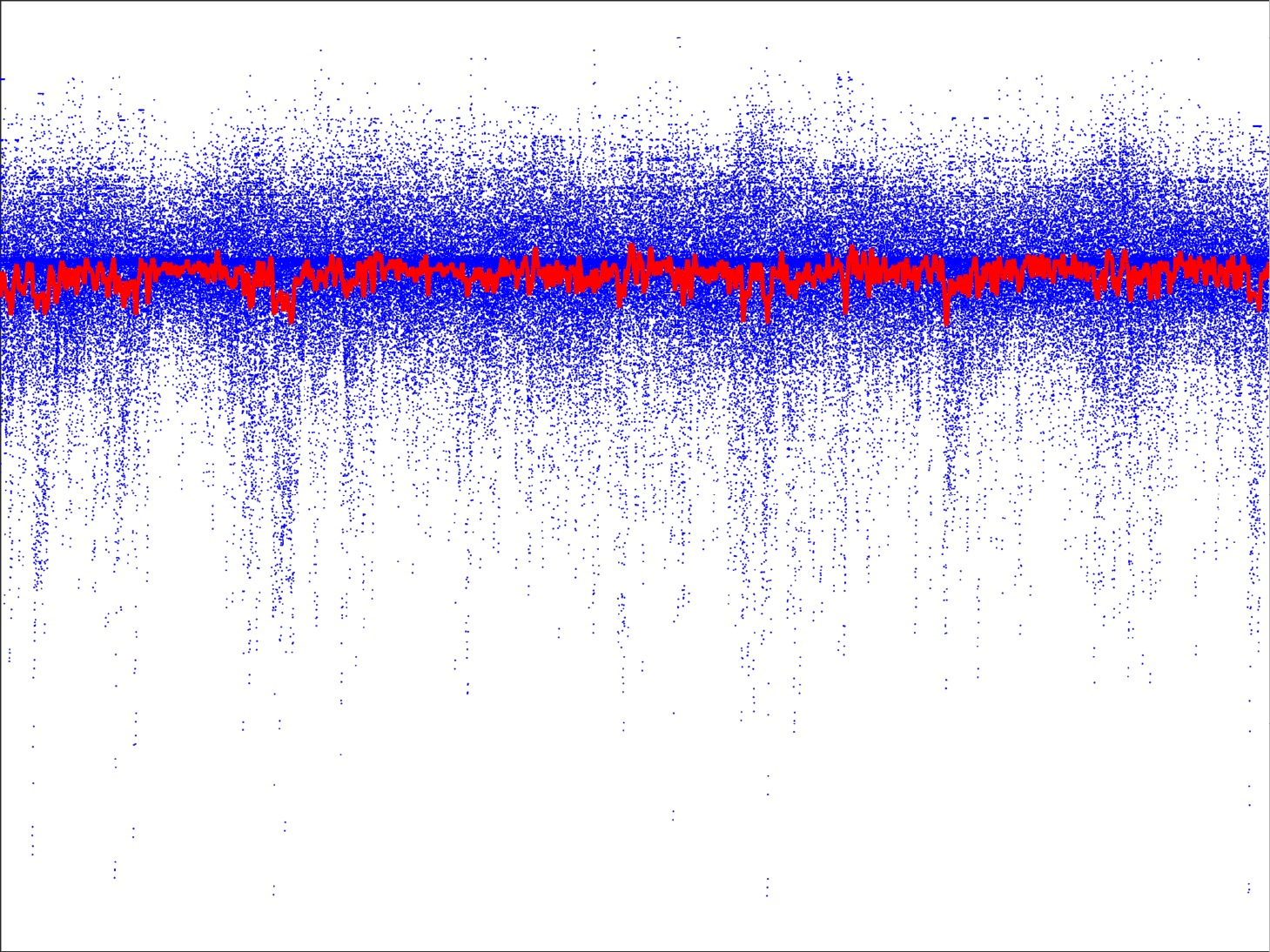};
\end{axis}
\end{tikzpicture}
    \caption{30\% CAV penetration.}
    \end{subfigure}
    
    \vskip\baselineskip
    
    \begin{subfigure}{0.48\textwidth}
    \hspace{1.0em}
    \begin{tikzpicture}
\begin{axis}[%
    xlabel style={font=\color{white!15!black}},
    xlabel={Time [min]},
    xtick={175,1250,2345,3415},
    xticklabels={{4},{21},{39},{56}},
    ylabel style={font=\color{white!15!black}},
    ylabel={Velocity [m/s]},
    enlargelimits=false, 
    axis on top,  
    width=6.5cm,
    legend style={
        draw=white!15!black,
        legend cell align=left,
        legend pos=south east}
    ]
    \addlegendimage{blue,mark=*}
    \addlegendentry[color=black]{Velocity Sample};
    \addlegendimage{red}
    \addlegendentry[color=black]{Average};
    ]
\addplot graphics [%
    xmin=107,
    xmax=3600,
    ymin=0,
    ymax=28] {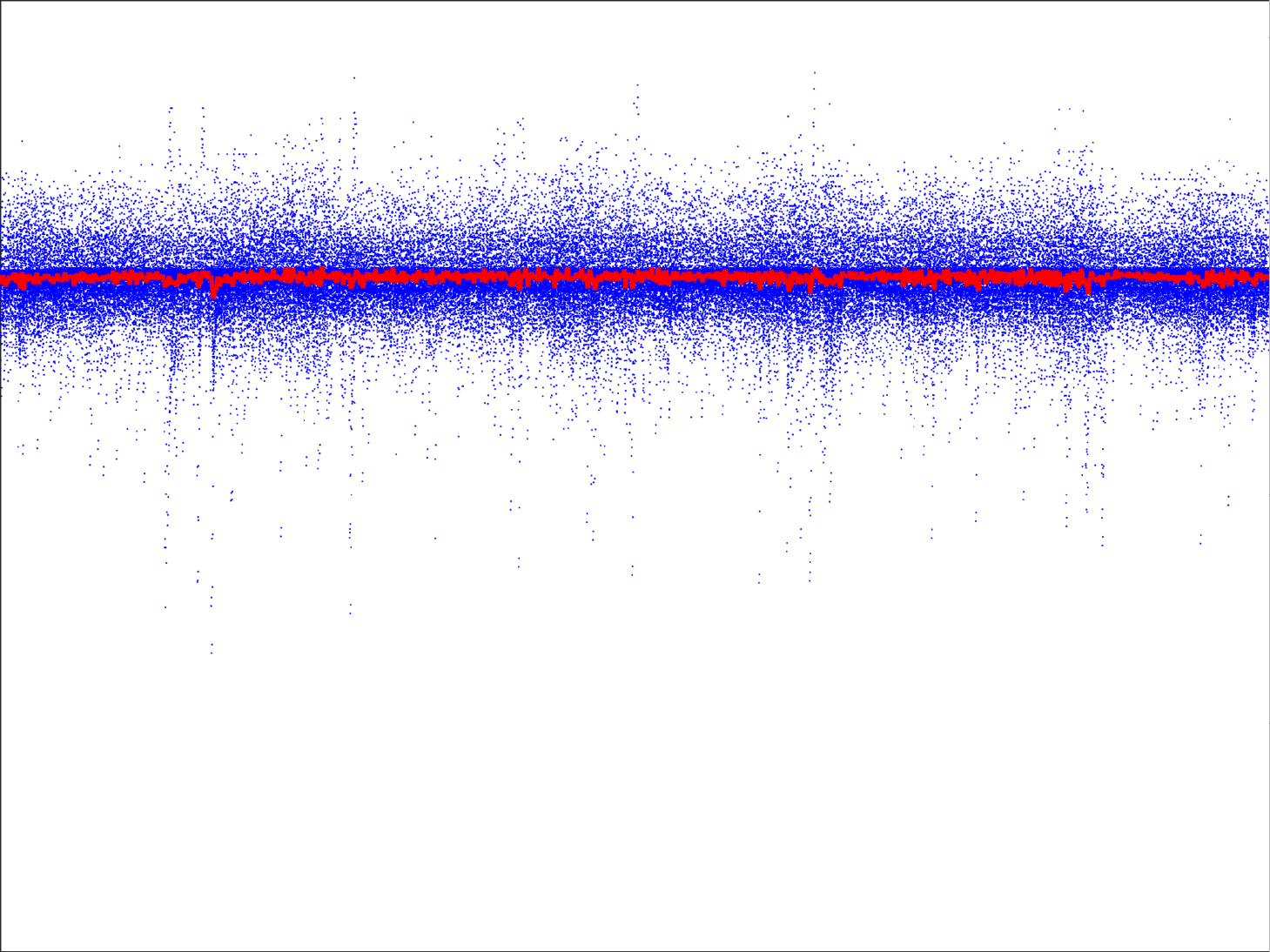};
\end{axis}
\end{tikzpicture}
    \caption{70\% CAV penetration.}
    \end{subfigure}
    \hfill
    \begin{subfigure}{0.48\textwidth}
    \hspace{1.0em}
    \begin{tikzpicture}
\begin{axis}[%
    xlabel style={font=\color{white!15!black}},
    xlabel={Time [min]},
    xtick={175,1250,2345,3415},
    xticklabels={{4},{21},{39},{56}},
    ylabel style={font=\color{white!15!black}},
    ylabel={Velocity [m/s]},
    enlargelimits=false, 
    axis on top,  
    width=6.5cm,
    legend style={
        draw=white!15!black,
        legend cell align=left,
        legend pos=south east}
    ]
    \addlegendimage{blue,mark=*}
    \addlegendentry[color=black]{Velocity Sample};
    \addlegendimage{red}
    \addlegendentry[color=black]{Average};
    ]
\addplot graphics [%
    xmin=107,
    xmax=3600,
    ymin=0,
    ymax=28] {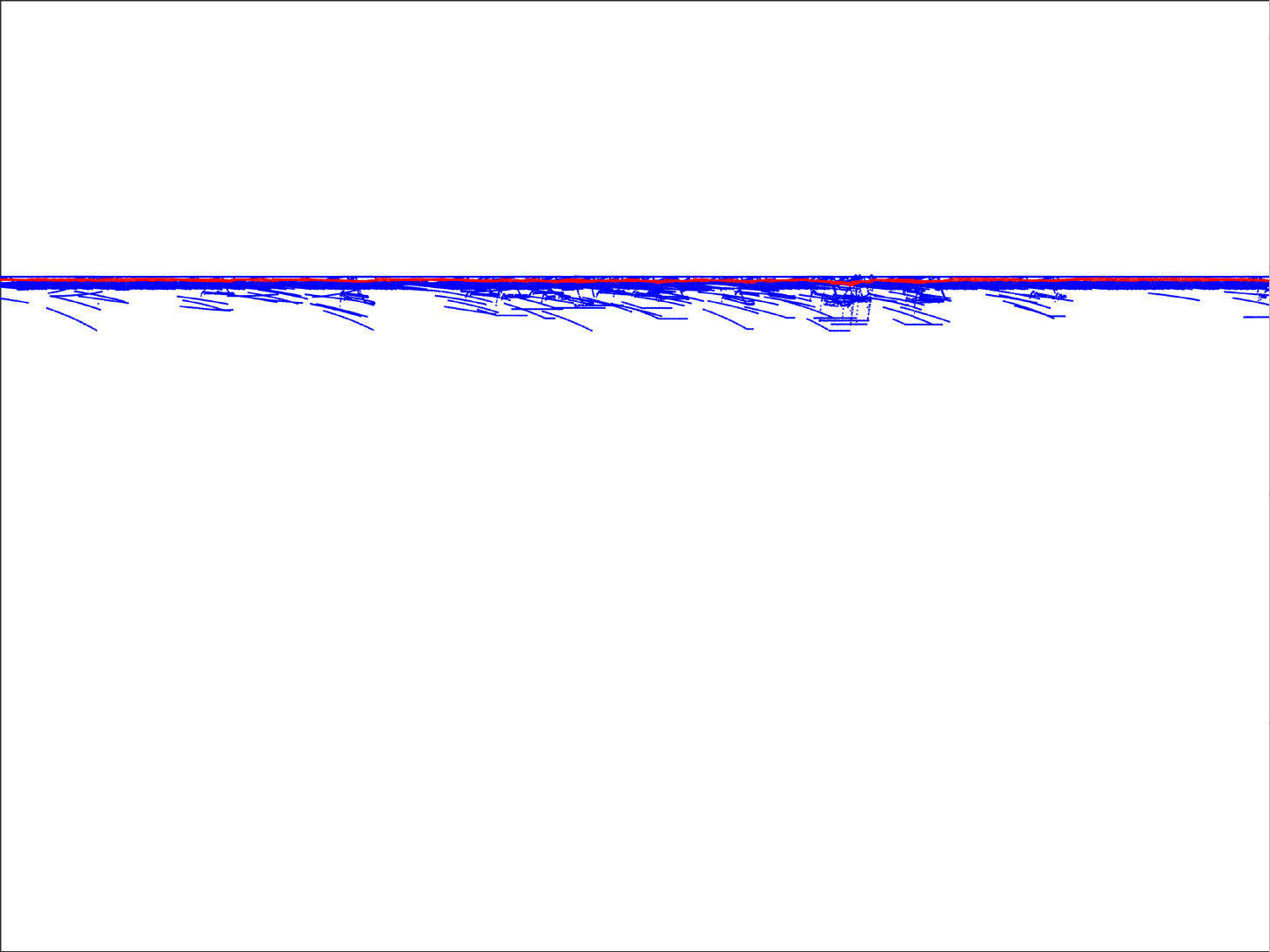};
\end{axis}
\end{tikzpicture}
    \caption{100\% CAV penetration.}
    \end{subfigure}
    
    \caption{Velocities of vehicles in the network at various CAV penetrations and 2000 veh/hour.}
    \label{fig:vel}
\end{figure}

In further support of this argument, we examine the velocity behavior of vehicles in the network. Figure \ref{fig:vel} depicts a scatter plot of all measured velocities, as well as the average velocity of all vehicles at each instance in time, for 2000 veh/hour. To again observe traffic smoothing, as CAV penetration increases, the velocities become more tightly packed to the average velocity profile. In addition, full stops that occurred for some vehicles no longer occur with increasing CAV penetration. With reductions in vehicle acceleration and velocity variance, the fuel economies improved.

%%%%%%%%%%%%%%%%%%%%%%%%%%%%%%%%%%%%%%%%%%%
%%% Conclusion                          %%%
%%%%%%%%%%%%%%%%%%%%%%%%%%%%%%%%%%%%%%%%%%%
\section{Conclusion and Future Work}\label{sec:Conclusion}

This paper studied the energy and traffic impact of a connected and anticipative car-following controller integrated into a PTV VISSIM microsimulation environment mixed with human-modeled drivers. In doing so, realistic VISSIM simulations were set up by utilizing time headways from real traffic data in a high-density highway scenario and imposing realistic acceleration capabilities of vehicles. Delays in the control algorithm due to computation and communication were also considered.

The proposed MPC algorithm utilized a cost function that tracks desired headway and penalizes acceleration, and constraints are introduced to model passenger vehicle capabilities, as well as enforce safety considerations through the use of chance constraints. A trade-off between traffic compactness and safety was balanced through the use of said chance constraints, which showed that traffic compactness was not compromised through the introduction of automated vehicles. With market penetrations of at least 30\% automated vehicles equipped with connectivity, road capacity was significantly improved. Further, a prediction model of constant acceleration is assumed for the preceding vehicle when anticipating trajectories in human vehicles.

It was found that the controller was effective in producing energy-efficient results. When evaluating a conventional powertrain, we found that MPC vehicles performed at 10\% higher energy efficiency over human drivers at low volume, and performed at 20\% higher energy efficiency over human drivers at high volume - while improving in capacity of the road over the all-human driver scenario. Secondary effects were seen in that traffic was smoothed to dissipate shockwaves, and so human driver fuel economy was additionally improved. Overall, up to 25\% higher fuel benefits in the entire fleet were observed in all-automated scenarios.

Further energy evaluation was conducted using electric and hybrid powertrains as well. Largely due to regenerative braking which improved human driver energy usage, energy improvements because of MPC presence in the fleet were not as pronounced, though we found up to 6\% improvement in energy efficiency over human drivers with electric vehicles, and we found up to 9\% improvement in energy efficiency over human drivers with hybrid vehicles.

This paper describes a framework that can be used to evaluate impact of autonomous vehicles, where methods depicted here can be applied to more sophisticated scenarios deserving of separate papers. 
In general, multi-lane controllers can be explored, whereas high fidelity VISSIM simulations of arterial and highway environments can be studied.

%%%%%%%%%%%%%%%%%%%%%%%%%%%%%%%%%%%%%%%%%%%

\section*{Acknowledgment}
This work was supported by an award from the U.S. Department of Energy Vehicle Technologies Office (Project DE-EE0008232).

\printcredits

\Urlmuskip=0mu plus 1mu\relax
\bibliographystyle{cas-model2-names}
\bibliography{references}

\end{document}